\documentstyle[preprint,aps,version2]{revtex}
\tightenlines

\newcommand{\be}{\begin{equation}}
\newcommand{\ee}{\end{equation}}
\newcommand{\beq}{\begin{eqalignno}}
\newcommand{\eeq}{\end{eqalignno}}
\newcommand{\epem}{\mbox{$e^+e^-$}}

\newcommand{\gev}{{\rm\,GeV}}
\newcommand{\tev}{{\rm\,TeV}}
\newcommand{\fb}{{\rm\,fb}}
\newcommand{\picob}{{\rm\,pb}}
\newcommand{\ifb}{{\rm\,fb}^{-1}}
\newcommand{\ipb}{{\rm\,pb}^{-1}}
\newcommand{\mz}{M_Z}
\newcommand{\mw}{M_W}
\newcommand{\cosb}{\cos\beta}
\newcommand{\sinb}{\sin\beta}
\newcommand{\tanb}{\tan\beta}
\newcommand{\cosw}{\cos\theta_W}
\newcommand{\sinw}{\sin\theta_W}
\newcommand{\cossqw}{\cos ^2 \theta_W}
\newcommand{\sinsqw}{\sin ^2 \theta_W}
\newcommand{\chc}{\tilde{\chi}^{\pm}}
\newcommand{\chargino}{\tilde{\chi}^{\pm}_1}
\newcommand{\mchargino}{m_{\tilde{\chi}^{\pm}_1}}
\newcommand{\chcp}{\tilde{\chi}^+}
\newcommand{\chcm}{\tilde{\chi}^-}
\newcommand{\chn}{\tilde{\chi}^0}
\newcommand{\LSP}{\chn_1}
\newcommand{\mLSP}{m_{\chn_1}}
\newcommand{\snu}{\tilde{\nu}}
\newcommand{\msnu}{m_{\tilde{\nu}}}

\newcommand{\mslep}{m_{\tilde{l}}}

\newcommand{\msq}{m_{\tilde{q}}}
\newcommand{\stot}{\sigma_{\rm total}}

\newcommand{\mtot}{{\cal M}^{\rm tot}}
\newcommand{\mprod}{{\cal M}^{\rm prod}}
\newcommand{\mdecay}{{\cal M}^{\rm decay}}
\newcommand{\mum}{(\mu , M_2)}

\newcommand{\mpt}{\rlap{$\not$}p_T}

\newcommand{\afbchi}{A_{FB}^{\chc}}
\newcommand{\afbjj}{A_{FB}^{jj}}
\newcommand{\rhoplus}{\rho_{\chargino}}
\newcommand{\rhozero}{\rho_{\LSP}}
\newcommand{\mjjmax}{m_{jj}^{\rm max}}
\newcommand{\Ejjmax}{E_{jj}^{\rm max}}
\newcommand{\Ejjmin}{E_{jj}^{\rm min}}
\newcommand{\Ejjm}{E_{jj}^0}
\newcommand{\sigh}{\sigma_{\rm hadronic}}
\newcommand{\sigl}{\sigma_{\rm leptonic}}
\newcommand{\sigm}{\sigma_{\rm mixed}}
\newcommand{\sigy}{\sigma_Y}
\newcommand{\sigi}{\sigma_i}
\newcommand{\ds}{\displaystyle}

\newcommand{\half}{\frac{1}{2}}
\newcommand{\third}{\frac{1}{3}}
\newcommand{\twothird}{\frac{2}{3}}
\newcommand{\fourth}{\frac{1}{4}}
\newcommand{\tanw}{\tan\theta_W}
\newcommand{\order}{{\cal O}}
\newcommand{\rarr}{\rightarrow}

\begin{document}
\draft

\pagestyle{empty}
\preprint{SLAC--PUB--6497}
\medskip
\preprint{RU--94--67}
\medskip
\preprint{hep-ph/9408359}
\medskip
\preprint{August 1994}
\medskip
\preprint{T/E}

\vspace{-0.1in}

\begin{title}
Determination of Fundamental Supersymmetry Parameters\\
from Chargino Production at LEP II
\end{title}

\author{
Jonathan L. Feng
\thanks{Work supported by the Department of Energy, contract
DE--AC03--76SF00515.}
\thanks{Work supported in part by an NSF Graduate Research Fellowship.}
}

\begin{instit}
Stanford Linear Accelerator Center\\
Stanford University, Stanford, California 94309
\end{instit}

\author{
Matthew J. Strassler
\thanks{Work supported in part by the Department of Energy, contract
DE--FG05--90ER40559.}
}

\begin{instit}
Department of Physics and Astronomy\\
Rutgers University, Piscataway, New Jersey 08855
\end{instit}


\begin{abstract}

\vspace{-0.5in}

If accessible at LEP II, chargino production is likely to be one of the
few available supersymmetric signals for many years.  We consider the
prospects for the determination of fundamental supersymmetry parameters
in such a scenario.  The study is complicated by the dependence of
observables on a large number of these parameters. We propose a
straightforward procedure for disentangling these dependences and
demonstrate its effectiveness by presenting a number of case studies at
representative points in parameter space. Working in the context of the
minimal supersymmetric standard model, we find that chargino production
by itself is a fairly sensitive probe of the supersymmetry-breaking
sector. For significant regions of parameter space, it is possible to
test the gaugino mass unification hypothesis and to measure the gaugino
contents of the charginos and neutralinos, thereby testing the
predictions of grand unification and the viability of the lightest
supersymmetric particle as a dark matter candidate. For much of the
parameter space, it is also possible to set limits on the mass of the
electron sneutrino, which provide a valuable guide for future particle
searches.

\bigskip

\centerline{(Submitted to Physical Review {\bf D})}

\end{abstract}


\newpage
\pagestyle{plain}

\narrowtext

\section{Introduction}
\label{sec:intro}

The gauge hierarchy problem has motivated many approaches to extending
the standard model, and among them, supersymmetry (SUSY) is one of the
most promising \cite{Reviews}. If SUSY is to provide a solution to this
problem, it must be broken at energies of order 1 TeV, and so the
masses of supersymmetric particles must lie at or below this energy.
Because such energies are within reach of existing accelerators or
those that are scheduled to operate in the near future, SUSY
phenomenology has attracted much attention in recent years.

In many of the supersymmetric models that have been explored, charginos
are the lightest observable supersymmetric particles, and for this
reason, chargino searches have been particularly well-studied
\cite{srchs,Chia,Btl1,Btl2,Aachen,Chen,Grvz2,Ahn,Grvz}. The current
lower bound on the mass of the lighter chargino is 45 GeV
\cite{PDG,ALEPH}.  Chargino discovery studies have shown that, with the
characteristics currently expected to be reached at LEP II, $\sqrt{s}$
= 175--190 GeV and a luminosity of 200--500 $\ipb /{\rm year}/{\rm
experiment}$ \cite{LEP200Overview}, the discovery reach will extend
nearly to the kinematic limit, significantly extending the accessible
region of parameter space.  It is also worth noting that there are at
present tentative but tantalizing hints of the possible existence of
light charginos from the measurement of $A_{LR}$ at SLC \cite{SLD,negS}
and measurements of $\Gamma(Z \rightarrow b\bar{b})$ at LEP and SLC
\cite{ZbbLEP,ZbbSLD,Don,Kolda}.

If charginos are discovered at LEP II, they will provide one of the
first direct signals of supersymmetry and may well prove to be the most
promising candidates for precision supersymmetry studies for many
years. Although neutralino production is likely to accompany chargino
production, it often suffers from a significantly smaller cross section
\cite{neutralino} and may be more difficult to separate from
backgrounds \cite{Aachen,Chen}.  It is therefore natural to ask what
information about the parameters of supersymmetry can be obtained from
the chargino signal alone.

In the case of linear \epem\ colliders with $\sqrt{s} = 500 \gev$, the
question of precision measurements of sparticle masses and underlying
SUSY parameters has been addressed in a number of studies
\cite{JLC,Fujii,Orito,Vandervelde,Tsukamoto,Kon,squark,LeNLC}. These
studies have shown that if a number of sparticles are light enough to
be produced, their masses can be determined to high accuracy, and the
sparticle spectrum can provide stringent tests of standard theoretical
assumptions.  If light charginos exist, thousands of them will be
produced at LEP II, and precision measurements might also be possible
there.  However, in addition to the difficulties present in all studies
of SUSY signals, such as unobservable particles in the final state and
a wealth of unknown parameters, chargino production at LEP II suffers
from other difficulties not present in the 500 GeV collider studies. In
particular, the background from $W$ pair production will have a
stronger overlap with the chargino signal, and beam polarization, an
important diagnostic in linear collider experiments, will not be
available. The aim of this paper is to determine to what extent these
difficulties can be overcome, and to explore the prospects for the
determination of fundamental SUSY parameters in different regions of
parameter space.

A previous study \cite{LeLEP} addressed this question with the
assumptions that charginos are lighter than $W$ bosons, that the vacuum
expectations values of the two Higgs fields are roughly equal, and, for
most of the analysis, that sneutrinos are either very heavy or very
light.  We will relax these assumptions so that we may determine to
what extent they may be tested. Our study will be conducted in the
context of the minimal supersymmetric standard model (MSSM) without
gaugino mass unification. We will find that for some regions of
parameter space, it is possible to test gaugino mass unification, a
general prediction of supersymmetric grand unified theories, and also
to place strong bounds on the mass of the electron sneutrino, which
would provide a valuable guide for designing future sparticle searches.
In addition, we will see that it is often possible to measure the
gaugino contents of the charginos and neutralinos, which have
implications for the viability of the lightest neutralino as a dark
matter candidate \cite{darkmatter}.

Formally, every observable defines a hypersurface in the space of SUSY
parameters, and determining the parameters simply consists of finding
the intersections of these hyperplanes. Practically, there are many
possible observables with varying degrees of dependence on the
fundamental parameters and different experimental uncertainties, and
reducing the allowed volume to a small region is at first sight far
from straightforward. The most effective way to extract the underlying
parameters from the data is to perform a binned maximum likelihood fit,
and ultimately, this is what should be done.  However, such a procedure
does not provide much physical understanding of the results, nor does
it provide a useful way to visualize how the underlying parameters are
constrained by specific measurements.  In this study, we will discuss
observables one by one in a way that gives a straightforward strategy
for disentangling their complicated dependences. In the process we will
show which parameters can be tightly constrained by chargino production
and which cannot.  We hope that this study will provide some general
understanding of the results one may hope to achieve.  Of course, if
light charginos are found, this picture will be considerably sharpened
by strategies tailored to the particular point in parameter space that
is realized in nature.

In Sec.~\ref{sec:susyparam} we briefly review the MSSM. We state our
assumptions about the MSSM and discuss the theoretical prejudices that
we hope to test through our approach. We then describe the region of
parameter space in which charginos can be produced at LEP II and
present the SUSY parameter space that we hope to constrain.  In
Sec.~\ref{sec:observables} we describe the salient aspects of chargino
events and discuss the observables that will be most central to our
analysis.  Sec.~\ref{sec:simulation} contains a description of the
event simulation and the cuts used in the case studies.  The case
studies themselves are presented in Sec.~\ref{sec:studies}, where we
describe the strategy that we will follow in systematically
constraining the SUSY parameters and then apply it to a number of
representative points in parameter space.  We conclude with some final
comments and a summary of our results in Sec.~\ref{sec:conc}.

\section{SUSY parameter space and chargino production}
\label{sec:susyparam}

\subsection{The Minimal Supersymmetric Standard Model}

Our analysis will be performed in the context of the MSSM
\cite{Reviews,GH}, the simplest extension of the standard model that
includes supersymmetry. In this subsection, we explain which
assumptions about the MSSM we will make, and we introduce the SUSY
parameters that we hope to constrain.

The MSSM includes the usual matter superfields and two Higgs doublet
superfields

\begin{equation}\label{Hfields}
\hat{H}_1=\left( \begin{array}{c} \hat{H}_1^0 \\
                                  \hat{H}_1^- \end{array} \right)
\hspace{.2in} {\rm and} \hspace{.2in}
\hat{H}_2=\left( \begin{array}{c} \hat{H}_2^+ \\
                                  \hat{H}_2^0 \end{array} \right),
\end{equation}
where $\hat{H}_1$ and $\hat{H}_2$ give masses to the isospin
$-\frac{1}{2}$ and $+\frac{1}{2}$ fields, respectively.  These two
superfields are coupled in the superpotential through the term $ - \mu
\epsilon_{ij}\hat{H}_1^i\hat{H}_2^j$, where $\mu$ is the supersymmetric
Higgs mass parameter.  The ratio of the two Higgs scalar vacuum
expectation values is defined to be $\tanb \equiv \langle H^0_2 \rangle
/ \langle H^0_1 \rangle$. Soft supersymmetry-breaking terms
\cite{Girardello} for scalars and gauginos are included in the MSSM with

\begin{equation}\label{vsoft}
V_{\rm soft} = \sum_i m_i^2 |\phi_i |^2
+ \frac{1}{2} \left\{ \Big[M_1\tilde{B}\tilde{B} + \sum_{j=1}^{3}
M_2\tilde{W}^j\tilde{W}^j + \sum_{k=1}^{8} M_3\tilde{g}^k
\tilde{g}^k\Big]
+ {\rm h.c.}\right\}
+ \big[A {\rm\, terms\,} \big],
\ee
where $i$ runs over all scalar multiplets.  The ``$A$ terms'' are cubic
scalar terms that couple the Higgs scalars to squarks and to sleptons
with coefficients $A_{u_i}$ $A_{d_i}$, and $A_{e_i}$, where $i$ is the
generation number.  Studies of the MSSM often assume a number of
relations among the mass parameters $m_i$ and $M_i$; the relatively
weak assumptions made in this study will be detailed shortly.

The charginos and neutralinos of the MSSM are the mass eigenstates that
result from the mixing of the electroweak gauginos $\tilde{B}$ and
$\tilde{W}^j$ with the Higgsinos.  The charged mass terms that appear
are

\be
(\psi ^-)^T {\bf M}_{\chc} \psi^+ + {\rm h.c.},
\ee
where $(\psi^{\pm})^T = (-i\tilde{W}^{\pm}, \tilde{H}^{\pm})$ and

\be\label{chamass}
{\bf M}_{\chc} = \left( \begin{array}{cc}
 M_2                    &\sqrt{2} \, \mw\sinb  \\
\sqrt{2} \, \mw\cosb   &\mu                    \end{array} \right).
\ee
The chargino mass eigenstates are $\tilde{\chi}^+_i = {\bf
V}_{ij}\psi^+_j$ and $\tilde{\chi}^-_i = {\bf U}_{ij}\psi^-_j$, where
the unitary matrices {\bf U} and {\bf V} are chosen to diagonalize
${\bf M}_{\chc}$.  Neutral mass terms may be written as

\be
\frac{1}{2} (\psi ^0)^T {\bf M}_{\chn} \psi^0 + {\rm h.c.},
\ee
 where $(\psi^0)^T = (-i\tilde{B},-i\tilde{W}^3, \tilde{H}^0_1,
\tilde{H}^0_2)$ and

\be\label{neumass}
{\bf M}_{\chn} =
 \left( \begin{array}{cccc}
M_1             &0              &-\mz\cosb\sinw &\mz\sinb\sinw  \\
0               &M_2            &\mz\cosb\cosw  &-\mz\sinb\cosw \\
-\mz\cosb\sinw  &\mz\cosb\cosw  &0              &-\mu           \\
\mz\sinb\sinw   &-\mz\sinb\cosw &-\mu           &0     \end{array}
\right).
\ee
The neutralino mass eigenstates are $\chn_i = {\bf N}_{ij}\psi^0_j$,
where {\bf N} diagonalizes ${\bf M}_{\chn}$.  In  order of increasing
mass the four neutralinos are labeled $\chn_1$, $\chn_2$, $\chn_3$, and
$\chn_4$, and the two charginos, similarly ordered, are $\chc_1$ and
$\chc_2$. From the mass matrices in Eqs.~(\ref{chamass}) and
(\ref{neumass}), it can be seen that in the limits $\tanb \rightarrow
0$ and $\tanb \rightarrow \infty$ there is an exact symmetry $\mu
\leftrightarrow -\mu$.

In the form outlined above, the MSSM contains many unknown parameters,
and typically a number of simplifying assumptions are made.  These
assumptions are usually based on grand unified theories or minimal
supergravity.  As we hope to test such theories, we choose less
restrictive and more phenomenological assumptions, which we list below.
Having stated our assumptions, we will then explain our reasons for
choosing them and explore their implications.

\subsection{Our Basic Assumptions}

In this study, we will make the following assumptions:

\noindent (a) R-parity is conserved.

\noindent (b) The lightest supersymmetric particle (LSP) is the
lightest neutralino, $\LSP$.

\noindent (c) Sleptons and squarks have masses beyond the kinematic
limit of LEP II.

\noindent (d) The gluino is heavier than the lighter chargino.

\noindent (e) The intergenerational mixing in the squark, slepton, and
quark sectors is small and may be neglected in our analysis.

\noindent (f) The four {\it left-handed} squarks of the first two
generations are nearly degenerate at low energy with mass $\msq$, as
are all six {\it left-handed} sleptons with mass $\mslep$:

\be\label{scalarassumption}
\begin{array}{rcl}
m_{\tilde{u}_L}\approx & m_{\tilde{d}_L} & \approx
m_{\tilde{c}_L}\approx m_{\tilde{s}_L} \approx  \msq \\
m_{\tilde{\nu}_{eL}}\approx m_{\tilde{e}_L} \approx &
m_{\tilde{\nu}_{\mu L}} & \approx m_{\tilde{\mu}_L} \approx
m_{\tilde{\nu}_{\tau L}}\approx m_{\tilde{\tau}_L}\approx  \mslep  .
\end{array}
\ee
As will be discussed at length below, chargino production and decay are
highly insensitive to the masses of all other scalars, and we may
therefore set all squark masses to $\msq$ and all slepton masses to
$\mslep$ without loss of generality.

\noindent (g) The gaugino masses $M_i$ and the parameters $\mu$ and
$\tanb$ may be taken to be real, so that CP violation plays no role in
chargino events.

\noindent (h) The one-loop corrections to particle masses, chargino
production, and chargino decay do not introduce any large new
dependences on fundamental SUSY parameters.

\noindent (i) The parameters $M_1$ and $M_2$ are independent, {\it
i.e.}, we do {\it not} assume gaugino mass unification, as we are
hoping to test this prediction of grand unified theories.

With the assumptions listed, our analysis is applicable to the bulk of
parameter space available for study at LEP II.  However, there are a
number of small regions in the allowed parameter space where the
physics is qualitatively different from the norm, and these will
require special treatment.  Since our principal aim is to explore the
most general properties of chargino production, we will not study these
exceptional regions, though their existence will be noted in the
remarks below.  In Sec.~\ref{subsec:otherobs} we will briefly discuss
ways in which the unusual physics present in these cases might be
detected.

Assumption (a) is commonly made in supersymmetry studies, as it
prevents protons from decaying too quickly.  Given R-parity
conservation, the LSP is stable and must be among the decay products of
any sparticle. The LSP must be uncolored and uncharged, and in many
models it is the lightest neutralino $\LSP$, as we have assumed in (b).

Because the LSP is very weakly interacting and unobservable in
detectors, the first potentially observable SUSY signal must include
the production of other light sparticles. As we would like to study the
question of how much information can be obtained from the chargino
signal alone, we will limit ourselves to models in which the sleptons
and squarks have masses beyond the kinematic limit of LEP II, as given
in (c). This assumption, along with the small cross sections for gluino
production at \epem\ colliders, implies that any reasonably large SUSY
signal at LEP II must involve either the lighter chargino $\chc_1$ or
the second lightest neutralino $\chn_2$. Although it would simplify our
analysis, we cannot assume that the second lightest neutralino is
heavier than the lighter chargino, since, as can be seen from
Eqs.~(\ref{chamass}) and (\ref{neumass}), $m_{\chn_2}$ is not
independent of $\mchargino$ and $\mLSP$. In fact, in the region of
parameter space in which chargino production is accessible to LEP II,
$\chc_1$ and $\chn_2$ are very roughly degenerate, with the mass
difference typically in the range $-10 \gev \alt m_{\chn_2} -
\mchargino \alt 20 \gev$.  When $m_{\chn_2} < \mchargino$, it is
possible for the lighter chargino to decay through a cascade decay, in
which it decays to a $\chn_2$, which in turn decays to an LSP.  If the
mass splitting $\mchargino - \mLSP$ is small, or direct decays to the
LSP are suppressed by small couplings, the branching fraction for
chargino cascade decays may be non-negligible.  However, as the
coincidence of these conditions occurs only in a small fraction of
parameter space, we will not consider these decays further.

Under our assumptions, charginos decay to three-body final states
consisting of an LSP and either two quarks or two leptons. The current
lower bound on $\mLSP$ is roughly 20 GeV \cite{PDG,ALEPH}, and so for
charginos produced at LEP II with $\sqrt{s}= 190\gev$, $\mchargino -
\mLSP < 75 \gev < \mw$.  Along with assumptions (a) -- (d), this
implies that the $W$ bosons, squarks and sleptons in chargino decays
must be virtual particles.  The tree-level relation $m_{H^{\pm}} > \mw$
implies the same for charged Higgs bosons.  If the tree-level $H^{\pm}$
mass were very low and its one-loop corrections were large and
negative, the two-body decay $\chargino\rarr H^{\pm}\LSP$ could occur,
but we will assume that this is not the case.

We may now display the production and decay diagrams of charginos.
Chargino production from \epem\ collisions is given by the three
processes in Fig.~\ref{fig:feynprod} and includes $s$-channel $\gamma$
and $Z$ diagrams and $t$-channel $\snu_e$ exchange. Charginos decay to
the LSP either hadronically through virtual $W$, squark, or $H^{\pm}$
diagrams,

\be\label{hadronic}
\chcp\rightarrow (\LSP {W^+}^*, {\tilde{d}}^* u, \bar{d} {\tilde{u}}^*,
\LSP {H^+}^*) \rightarrow \LSP \bar{d} u,
\ee
where $u$ ($d$) represents an up- (down-) type quark, or leptonically
through virtual $W$, slepton, or $H^{\pm}$ diagrams,

\be\label{leptonic}
\chcp\rightarrow (\LSP {W^+}^*, {\tilde{l}}^* \nu, \bar{l}
{\tilde{\nu}}^*, \LSP {H^+}^* )\rightarrow \LSP \bar{l} \nu .
\ee
These decays are shown in Fig.~\ref{fig:feyndecay}. As each chiral
fermion has its own complex scalar partner, charginos decay
hadronically through six channels and leptonically through five (since
there is no right-handed sneutrino).

Charginos are too short-lived to be directly observed, so we must infer
everything from their decay products.  It will be convenient to refer
to the different types of chargino events by their decay modes.
However, since a $\tau$ lepton produced in the leptonic decay of a
chargino may itself decay either hadronically or leptonically, we must
distinguish between the particles that are directly produced at the
chargino decay vertices and those that are actually observed in the
detector.  To be precise, we establish the following terminology.  If
both charginos decay through the hadronic diagrams of
Fig.~\ref{fig:feyndecay}, we will call the event a ``hadronic mode''
event.  Events where both charginos decay through the leptonic diagrams
of Fig.~\ref{fig:feyndecay} will be called ``leptonic mode'' events,
and those where one chargino decays through a leptonic diagram and one
through a hadronic diagram will be called ``mixed mode'' events. On the
other hand, if we wish to group chargino events by their observed final
state, we will explicitly refer to the final state partons, using the
notation $2l$, $2j+l$, and $4j$ for two lepton, dijet plus lepton, and
four jet final states, respectively. In our notation, we will denote
the final state of a hadronically decaying $\tau$ lepton as $jj$. For
example, $2j+l$ events will include leptonic mode events in which one
chargino decays to a $\tau$ that decays hadronically and the other
chargino decays to a $\bar{\tau}$ that decays leptonically.

We will also need to identify the subset of $2j+l$ events that are
mixed mode events, that is, the $2j+l$ events in which the hadrons do
{\it not} come from a $\tau$.  These events will be called ``$Y$ mode''
events, the ``$Y$'' representing the topology of the lepton track and
the two jets. Hadronically decaying $\tau$ leptons produce a collimated
hadronic system of low invariant mass and, often, just a single charged
prong. In contrast, we will see that few dijet systems from chargino
decays have a low invariant mass, and so it is usually possible to
separate $Y$ events from the $2j+l$ events resulting from hadronic
$\tau$ decays.  As will be discussed below, lepton universality (which
follows from assumptions (e) and (f)) implies that a measurement of the
number of $Y$ mode events can be directly converted to a measurement of
the number of mixed mode events.

Before evaluating the importance of the various chargino decay
diagrams, we digress slightly to consider the possibility of studying
other SUSY signals. We have assumed that sleptons and squarks are
beyond the kinematic limit of LEP II. If sleptons and squarks are
within reach, they clearly will also give valuable information and will
improve the results we obtain here.  (Of course, if sleptons or squarks
are not only within reach of LEP II, but are also less massive than the
lighter chargino, the chargino will decay to two-body states containing
these particles, and a modified analysis will be necessary.) It is more
important to consider neutralino production. As noted above, $\chc_1$
and $\chn_2$ are very roughly degenerate, and therefore, if it is
possible to produce $\chcp_1 \chcm_1$ chargino pairs, it is likely that
$\chn_2$ pairs can be produced, and, in almost all cases, production of
$\chn_1 \chn_2$ is kinematically allowed.  If $M_2 \sim |\mu| \sim
\mw$, $\chn_3$ or even $\chn_4$ production may be possible at LEP II.
In principle neutralino pair production should also provide valuable
information. However, neutralino production cross sections are
typically significantly smaller than those for charginos and may differ
by as much as an order of magnitude or more in some regions of
parameter space \cite{Aachen,Chen,neutralino}. Furthermore, studies of
$\chn_1 \chn_2$ production at $\sqrt{s} = 190 \gev$ have concluded that
the signal suffers from a large background from $WW$ production in both
hadronic and leptonic modes \cite{Aachen,Chen}. For these reasons we
will not consider neutralino events further, other than to discuss
their impact on our ability to isolate the chargino signal. Insofar as
additional information about SUSY parameters can be obtained from
slepton, squark, and neutralino signals at LEP II, the results of our
study may be considered conservative.

{}From Figs.~\ref{fig:feynprod} and \ref{fig:feyndecay}, it is clear
that the chargino production and decay processes have a complex
dependence on the various scalar masses, with the sneutrino mass
$\msnu$ entering the production process, and all slepton and squark
masses entering the decay. In many versions of the MSSM, the slepton
and squark masses are assumed to be unified at a high energy scale.
When they are run down to low energies, typically the squarks acquire a
greater mass than the sleptons through their QCD interactions. In
principle, one would like to test this assumption. Unfortunately,
without some simplifications, the large number of independent squark
and slepton masses quickly complicates the analysis. We will make some
simple assumptions to bring this dependence under control.

It is first important to note that with the assumption of negligible
intergenerational mixings (assumption (e)), chargino events at LEP II
are highly insensitive to certain scalar masses, namely, those of the
third generation squarks, the right-handed squarks and sleptons, and
the charged Higgs boson.  To see this, we must discuss the scalar mass
spectrum in greater detail. Sfermion masses are given by the following
matrices:

\be\label{sup}
{\bf M}_{\tilde{u}}^2 =
 \left( \begin{array}{cc}
m_{\tilde{Q}}^2+m_u^2+\mz^2(\half-\twothird\sinsqw )\cos 2\beta
&m_u(A_u-\mu\cot\beta)  \\
m_u(A_u-\mu\cot\beta)
&m_{\tilde{U}}^2+m_u^2+\mz^2(\twothird\sinsqw )\cos 2\beta
\end{array} \right) \ ,
\ee
\be\label{sdown}
{\bf M}_{\tilde{d}}^2 =
 \left( \begin{array}{cc}
m_{\tilde{Q}}^2+m_d^2-\mz^2(\half-\third\sinsqw )\cos 2\beta
&m_d(A_d-\mu\tanb)  \\
m_d(A_d-\mu\tanb)
&m_{\tilde{D}}^2+m_d^2-\mz^2(\third\sinsqw )\cos 2\beta
\end{array} \right) \ ,
\ee
\be\label{selectron}
{\bf M}_{\tilde{e}}^2 =
 \left( \begin{array}{cc}
m_{\tilde{L}}^2+m_e^2-\mz^2(\half-\sinsqw )\cos 2\beta
&m_e(A_e-\mu\tanb)  \\
m_e(A_e-\mu\tanb)
&m_{\tilde{E}}^2+m_e^2-\mz^2(\sinsqw )\cos 2\beta
\end{array} \right) \ ,
\ee
\be\label{sneu}
{\bf M}_{\tilde{\nu}}^2 = m_{\tilde{L}}^2+\half\mz^2\cos 2\beta  \ ,
\ee
where the two-by-two matrices are in the basis $(\tilde{f}_L,
\tilde{f}_R)$. The masses $m_{\tilde{Q}}$, $m_{\tilde{U}}$,
$m_{\tilde{D}}$, $m_{\tilde{L}}$, and $m_{\tilde{E}}$ are the soft
SUSY-breaking scalar masses of Eq.~(\ref{vsoft}), and the $A_i$ are the
coefficients of the SUSY-breaking cubic scalar terms. The mass matrices
of the other two generations are identical in form.  In the top and
bottom squark sectors, the off-diagonal terms of the mass matrices can
be large, leading to large left-right mass splittings and light top and
bottom squarks.  However, because the charginos we are studying are
lighter than the top quark, and because we have assumed that
intergenerational mixings are negligible, decays of charginos through
third generation squarks are heavily suppressed. Thus, peculiarities of
the third generation are irrelevant for our analysis. For all other
squarks and sleptons, left-right mixings are usually negligible and the
masses of the sparticles are given by the diagonal elements of the
matrices. An exception occurs when $m_{\tau}(A_{\tau} - \mu\tanb ) \sim
m_{\tilde{L}}^2$, and similarly for muons and strange quarks, which is
possible in certain corners of SUSY parameter space (see
Sec.~\ref{subsec:boundsofparams}).  However, we will ignore left-right
mixing in this study and merely note in Sec.~\ref{subsec:otherobs} that
it would have observable consequences.

Let us now consider the right-handed scalars of the three slepton and
first two squark generations.  Right-handed scalars couple only to the
Higgsino component of the chargino.  As these couplings are the
supersymmetric analogues of Higgs couplings, they are suppressed by
either $m_u/(M_W \sin\beta)$ or $m_d/(M_W \cos\beta)$, where $m_u$ and
$m_d$ are the masses of the corresponding standard model fermion.  Such
couplings are important only for extreme values of $\tanb$, and so for
almost all of parameter space, chargino events are insensitive to the
masses of right-handed scalars.  The $H^{\pm}$ amplitude is suppressed
by similar couplings, and may also be ignored.

Thus, only the scalar masses listed in Eq.~(\ref{scalarassumption}) are
relevant. Left-handed scalars couple to charginos through their gaugino
components, and these couplings are not suppressed. However, in any
given generation, the left-handed squarks (sleptons) belong to the same
SU(2) doublet, and so have the same soft SUSY-breaking mass term
$m_{\tilde{Q}}$ ($m_{\tilde{L}}$).  Their masses are therefore split
only by the last terms of the diagonal entries of the scalar mass
matrices, the $D$ terms, which induce a mass splitting that is
typically of order 20 GeV or less for the masses we will consider. Such
splittings are not important for this study, and so, within each
generation, the left-handed squarks and sleptons may be taken to be
roughly degenerate. Furthermore, there are bounds on slepton and squark
non-degeneracy between {\it different} generations from $\mu
\rightarrow e \gamma$ and flavor changing neutral current constraints
\cite{FCNC}. Motivated by these considerations, we adopt assumption
(f).  Because the scalar masses that are not constrained in assumption
(f) are irrelevant for chargino events at LEP II, we may set them to
any reasonable values.  For convenience only, we will assume throughout
our study that {\it all} sleptons have the same mass $\mslep=\msnu$,
and similarly that {\it all} squarks have mass $\msq$; it should be
remembered, however, that our results depend only on the weaker
assumption (f).

Throughout this study, we will assume that one-loop corrections do not
greatly affect our analysis (assumption (h)).  Studies have shown that
one-loop effects on chargino and neutralino masses are generically only
a few percent \cite{Pierce,Lahanas}, and so we do not expect this
assumption to be very restrictive.  Even if one-loop corrections are
substantial, as long as they do not introduce qualitatively new
dependences on SUSY parameters into the observables we use, the
analysis presented in this study will still be applicable without large
modifications.

Often gauge coupling constants and gaugino masses are assumed to unify
at some high scale.  This assumption implies that even at lower energy
scales we have (to the accuracy of one-loop renormalization group
equations) \cite{Inoue}

\be\label{gaugunif}
\frac{M_2}{g^2_2} = \frac{3}{5} \frac{M_1}{g^2_1} = \frac{M_3}{g^2_3},
\ee
or approximately $M_1\approx \frac{1}{2} M_2$ and $M_3\approx
\frac{10}{3} M_2$ at $M_Z$. We may ignore the gluino mass $M_3$, since,
given assumption (d), gluinos enter chargino production and decay only
through loop diagrams, which are likely to be small.  As noted
previously, since one of our main goals is to test the unification of
$M_1$ and $M_2$, we will avoid assuming a universal gaugino mass and
will take these parameters to be independent (assumption (i)).  It is
possible without loss of generality to set $M_2 \ge 0$, and we will
follow this convention. Without the gaugino mass unification
assumption, however, $M_1$ may be either positive or negative.

\subsection{Regions of Parameter Space}

Given the discussion above, the SUSY parameter space of the MSSM that
is relevant to our study of chargino production is given by the six
parameters $(\mu, M_2, \tanb, M_1, \mslep, \msq)$. We can now examine
the regions of parameter space for which chargino production is
kinematically allowed at LEP II with $\sqrt{s}=190\gev$. The chargino
mass $\mchargino$ is completely determined by the three parameters
$\mu$, $M_2$, and $\tanb$.  In Fig.~\ref{fig:chargcont}, contours of
constant $\mchargino$ in the $\mum$ plane are plotted for fixed
$\tanb=4$; the contours are similar for other values of $\tanb$.  The
cross-hatched regions along the $M_2=0$ and $\mu=0$ axes are
experimentally excluded by lower limits on sparticle masses
\cite{PDG,ALEPH}, and the hatched region is the inaccessible region
where $\mchargino > 95 \gev$.  In Fig.~\ref{fig:neutcont} we plot
constant $\mLSP$ contours for $\tanb=4$ and, since $\mLSP$ depends on
$M_1$, three different values of $M_1/M_2$. (For these plots, only the
experimental limits $\mLSP > 20 \gev$ and $\mchargino > 45 \gev$ have
been included, as the other chargino and neutralino mass limits assume
gaugino mass unification \cite{PDG,ALEPH}.  Inclusion of the mass
bounds for the other neutralinos and chargino in the case
$M_1=\frac{1}{2} M_2$ extends the excluded region only slightly. The
exact shape of the experimentally excluded region will be unimportant
for this study.)

It is convenient to further divide the $\mum$ plane into regions based
on the gaugino contents of the light gauginos.  To quantify this, we
define the gaugino contents of the lighter chargino and LSP to be
\cite{darkmatter}

\be
\begin{array}{rcl}
\rhoplus &\equiv& |{\bf V}_{11}|^2 \\
\rhozero &\equiv& |{\bf N}_{11}|^2 + |{\bf N}_{12}|^2 \ .
\end{array}
\ee
(We have arbitrarily chosen to define $\rhoplus$ in terms of ${\bf
V}_{11}$ instead of ${\bf U}_{11}$.  These differ little throughout
parameter space, and for the purposes of defining $\rhoplus$, the
discrepancy is not important.)  The variables $\rhoplus$ and $\rhozero$
lie in the range $0 \le \rhoplus,\rhozero \le 1$; $\rhoplus$
($\rhozero$) is zero when $\chcp_1$ ($\LSP$) is pure Higgsino and is
one when $\chcp_1$ ($\LSP$) is pure gaugino.  Although they may differ
substantially in certain regions of parameter space, $\rhoplus$ and
$\rhozero$ are correlated: when $M_2,|M_1| \ll |\mu|$, both $\chargino$
and $\LSP$ are essentially gaugino states, while in the opposite limit,
$M_2,|M_1| \gg |\mu|$, they are both dominated by their Higgsino
components. We will present results for both quantities, and will find
that the bounds we obtain for them are roughly the same. The quantity
$\rhozero$ has implications for the viability of the LSP as a dark
matter candidate. As shown in a number of studies \cite{darkmatter},
$\chn_1$ is a good dark matter candidate when it is gaugino-like with
$\rhozero \agt 0.9$. For $\rhozero \alt 0.9$, the LSPs annihilate so
quickly in the early universe that they provide insufficient mass today
to be interesting dark matter candidates.  The gaugino content
$\rhozero$ is therefore an important parameter for us to determine.
However, because $\rhoplus$ depends only on the three parameters $\mu$,
$M_2$, and $\tanb$, it is the more convenient of the two quantities to
use to divide the parameter space. Contours of equal $\rhoplus$ are
plotted in Fig.~\ref{fig:rhocont}. Although the specific boundaries are
not particularly important, for definiteness we will refer to the
region with $\rhoplus \ge 0.9$ as the gaugino region, the region with
$\rhoplus \le 0.2$ as the Higgsino region, and the region with $0.2 <
\rhoplus < 0.9$ as the mixed region.  With these definitions, roughly
speaking, $\chn_1$ is a good dark matter candidate in the gaugino
region, but is not a viable candidate in the mixed and Higgsino regions.

It is evident from Figs.~\ref{fig:chargcont}, \ref{fig:neutcont}, and
\ref{fig:rhocont} that accurate determinations of $\mchargino$ and
$\mLSP$ are not enough to determine the gaugino content $\rhoplus$.
While a measurement of $\mLSP \approx \frac{1}{2} \mchargino$ might be
taken as evidence that the SUSY parameters lie in the gaugino region
and that gaugino masses unify, this is not the only possibility.  For
example, the masses $\mchargino \approx 80 \gev$ and $\mLSP \approx 40
\gev$ can be obtained with the parameters $(\mu,M_2,\tanb,M_1/M_2) =
(-400,75,4,0.5)$ in the gaugino region with $\rhoplus = 1.00$, and also
with the parameters $(-78,170,4,0.25)$ in the mixed region with
$\rhoplus = 0.34$. As will be seen below, more careful analysis can
differentiate between such possibilities.

We have now found the regions in which chargino production is allowed.
However, if the splitting between the masses of the chargino and the
LSP is very small, and the charginos are produced with low velocity,
the chargino decay products will have very low energy in the laboratory
frame and may be too soft to be experimentally useful. The approximate
relations \cite{GH2}

\be\label{thumb}
\mLSP \approx \min\{|\mu|, M_2, |M_1|\} \quad {\rm and} \quad
\mchargino \approx \min\{|\mu|, M_2\}
\ee
are valid in the far Higgsino and far gaugino regions.  With these in
mind, it is easy to see that for increasing $M_2$ and fixed $M_1/M_2$
in the Higgsino region, $\chargino$ and $\LSP$ become more and more
degenerate.  The maximum and minimum energies for dijet or $l\nu$
systems from chargino events are

\be\label{emaxmin}
E^{\rm max, min} = \frac{E_b}{2} \left[1\pm
\left(1-\frac{\mchargino^2}{E_b^2}\right)^{\frac{1}{2}}\right]
\left[1-\frac{\mLSP ^2}{\mchargino ^2} \right],
\ee
where $E_b$ is the beam energy.  The maximum energy for single jets and
leptons is also given by $E^{\rm max}$, which is plotted in
Fig.~\ref{fig:emax} for the case $M_1/M_2=0.5, \tanb=4$.  We see that,
in this case, the far Higgsino region with $M_2 \agt 500 \gev$,
contains areas in which all decay products have energies $\alt 10
\gev$. Thus, in these regions chargino production may be visible but
difficult to use in precision studies. As $M_2$ increases further, it
becomes difficult even to detect the chargino signal above background.
The problem of soft decay products is generic only in the far Higgsino
region.  In all other regions within the bands shown in
Fig.~\ref{fig:chargcont}, as long as $\mchargino$ lies somewhat below
the beam energy, and the splitting between $\mchargino$ and $\mLSP$ is
not anomalously small, chargino production can be observed and studied
at LEP II.

For extremely large values of $M_2$, the lifetime of the chargino
becomes long, and one might hope to tag charginos by looking for tracks
which do not intersect the interaction point. From the formula for the
chargino decay width presented later in Eq.~(\ref{DecayWidthA}), one
may estimate the chargino decay length, which is of order the impact
parameter for these tracks. Roughly, for charginos to travel 10 (100)
$\mu$m before decaying requires $M_2\agt 2 (3) \tev$.  Such large
values of $M_2$ are disfavored by fine-tuning considerations, as
discussed below.

\subsection{Boundaries of Parameter Space}
\label{subsec:boundsofparams}

In this subsection, we specify the boundary of the region of parameter
space that we will investigate.  The six SUSY parameters may be
restricted on the basis of fine-tuning prejudices and other
considerations. We discuss each parameter in turn.

As noted in Sec.~\ref{sec:intro}, if SUSY is to naturally explain the
electroweak scale, the SUSY-breaking masses $M_1$ and $M_2$ must be
less than or of order 1 TeV. In fact, if charginos are discovered, the
parameter $M_1$ may be further bounded.  Eq.~(\ref{thumb}) implies that
in the far gaugino and far Higgsino regions, if $M_1 \agt M_2$,
$\chargino$ and $\LSP$ are virtually degenerate.  Thus, the condition
that $\mchargino - m_{\LSP}$ be large enough that the decay products
are detectable implies to a good approximation that $|M_1| < M_2$, and
so we will also impose this constraint. Although $\mu$ is not a
SUSY-breaking parameter, fine-tuning considerations constrain it to lie
at or below the TeV scale.

The parameters $\mslep$ and $\msq$ are also SUSY-breaking masses, and
therefore must also be less than or of order 1 TeV.  Sleptons and
squarks with masses of 1 TeV are effectively decoupled and are
indistinguishable from those with infinite mass. We therefore limit the
analysis to $\mslep , \msq \le 1 \tev$. As we are considering the
scenario in which only charginos are produced at LEP II, we take
$\mslep \ge 100 \gev$. The squark mass lower bound from hadron
colliders is likely to be approximately 150 GeV when LEP II begins
operation, and we therefore take this as the lower bound on $\msq$.

The quantity $\tanb$ is more difficult to delimit. If one assumes the
desert hypothesis, applies the MSSM renormalization group equations to
the Higgs-fermion Yukawa couplings, and demands that the couplings
remain finite up to a scale $\Lambda = 10^{16} \gev$, one finds that
for the present top quark mass measurement of $m_t = 174 \pm
10^{+13}_{-12} \gev$ \cite{toplimit}, the bounds on $\tanb$ are $1 \alt
\tanb \alt 50$. We will adopt these limits.

In summary, given the assumptions above, our task is to explore and
restrict the six-dimensional SUSY parameter space given by

\be\label{bounds}
\begin{array}{rcccl}
-1 \tev & \alt & \mu   & \alt & 1 \tev \\
0       & \le  & M_2   & \alt & 1 \tev \\
1       & \le  & \tanb & \le  & 50     \\
-M_2    & \le  & M_1   & \le  & M_2    \\
100 \gev& \le  &\mslep & \le  & 1 \tev \\
150 \gev& \le  &\msq   & \le  & 1 \tev \ .
\end{array}
\ee

\section{Observables of Chargino Production}
\label{sec:observables}

In this section, we will discuss the observables that we will use to
restrict SUSY parameter space. As stated earlier, our goal is to gain
an understanding of chargino pair production by using as much analytic
information as possible. For this reason we will not study observables
for which no analytic formulae can easily be found, such as the
distribution of lepton energies in $2j+l$ events, even though these
quantities contain information which is not accessible through the
observables we consider. It is therefore probable that a global
likelihood fit to the data will be able to put tighter bounds on
supersymmetry parameter space than we will claim below. In this sense,
our results are conservative.

The four quantities that will be central to our analysis are the
chargino and neutralino masses, the total cross section for chargino
production, and the leptonic branching fraction:

\be\label{fourobs}
\begin{array}{l}
\mchargino(\mu, M_2, \tanb) \ , \\
\mLSP(\mu, M_2, \tanb, M_1) \ , \\
\stot(\mu, M_2, \tanb, \msnu=\mslep) \ , \\
B_l (\mu, M_2, \tanb, M_1, \mslep, \msq)\equiv
\frac{\displaystyle \Gamma(\chcp \rightarrow \chn \bar{l} \nu)}
{\displaystyle \Gamma(\chcp \rightarrow {\rm anything})} \ .
\end{array}
\ee
Of course, four observables will not allow us to determine six
parameters, but, as will be seen in Sec.~\ref{sec:studies}, these four
observables can often restrict the parameter space to a region in which
the quantities of greatest interest are already somewhat constrained.
The forward-backward asymmetry of chargino production $\afbchi$ will
also be discussed, but for a number of reasons to be mentioned below,
we will not use this quantity directly. No other variables were found
that could be studied without performing Monte Carlo simulations at a
large number of points in parameter space. The left-right asymmetry in
the production cross section requires polarized electron beams and is
inaccessible at LEP II, but has implications for chargino production at
threshold and will also be discussed below.

In the following subsections, we will consider each observable, first
analyzing its dependence on the underlying SUSY parameters, and then
discussing the method by which it may be extracted from chargino event
samples. This section will be confined to theoretical considerations;
experimental issues will be discussed in Sec.~\ref{sec:simulation}. In
particular, discussion of issues involving experimental efficiencies
and minor subtleties involving the hadronic decays of the $\tau$ lepton
will be deferred to the following sections. In Sec.~\ref{sec:studies},
the measurements suggested in this section will be applied to Monte
Carlo simulation case studies at specific points in parameter space,
and results will be obtained with cuts, finite detector resolution, and
finite statistics included.

\subsection{Chargino and Neutralino Masses}
\label{subsec:masses}

The chargino mass $\mchargino (\mu, M_2, \tanb)$ and the LSP mass
$\mLSP (\mu, M_2, \tanb, M_1)$ are sensitive to fundamental parameters
of supersymmetry and are relatively easy to determine. Their
dependences on the underlying SUSY parameters were discussed in
Sec.~\ref{sec:susyparam}.  Here we note only that in the gaugino region
$\mchargino \approx M_2$ and $\mLSP \approx |M_1|$, and the masses are
therefore directly sensitive to two fundamental parameters. In
contrast,  both masses are close to $|\mu|$ in the Higgsino region
(unless $|M_1|<|\mu|$, in which case the LSP can have a mass near
$|M_1|$) and the mass splitting $\mchargino - \mLSP$ is a complicated
function of several parameters.

The masses $\mchargino$ and $\mLSP$ can be measured in chargino events
in at least two ways.  It is impossible to kinematically reconstruct
chargino pair production events, since the charginos' decay products
include two unobservable LSPs.  However, because the unobserved LSPs
are typically quite massive and carry off large energies and momenta,
chargino events with two jets and an isolated lepton are easily
separated from standard model backgrounds by a series of cuts, as we
will see in Sec.~\ref{sec:simulation}. After imposing such cuts, one
can find the dijet energy $E_{jj}$ and dijet mass $m_{jj}$ for each of
the remaining events.  The endpoints of the dijet energy and mass
spectra are completely determined by $\mchargino$ and $\mLSP$, with the
endpoints of the $E_{jj}$ spectrum given by Eq.~(\ref{emaxmin}), and
the $m_{jj}$ distribution lying between zero and $\mchargino - \mLSP$.
If at least two of the three endpoints are sufficiently sharp to be
well-measured, they can be used to precisely determine $\mchargino$ and
$\mLSP$.  Of course, detector and beam effects will smear the
endpoints, but in Sec.~\ref{sec:studies} we will see that the masses
may still be measured to a few GeV with this method.

An energy scan at the chargino production threshold provides an
alternate determination of $\mchargino$ \cite{LeLEP}. In
Fig.~\ref{fig:threshold}, the total cross section as a function of
$\sqrt{s}$ (solid curve) is plotted for the particular point in
parameter space $(\mu, M_2, \tanb, \msnu) = (-400, 75, 4, 200)$.  For
comparison, a unit of R is also given (dashed curve).  The sudden rise
in cross section, characteristic of fermion production, makes possible
a highly accurate determination of $\mchargino$. Such behavior is
common for all points in parameter space. Near threshold the charginos
are nearly at rest, so the only unknown momentum in the decay
$\chargino\rarr q\bar{q} \chn_1$ is that of the LSP.  By reconstructing
the hadronically decaying chargino in a mixed mode event, one obtains
the mass of the LSP.  Although this method is likely to provide a
significantly more accurate determination of the chargino mass
\cite{LeLEP}, we will not assume in our case studies that an energy
scan will be performed, instead relying solely on the endpoint
determinations.

\subsection{Differential Cross Section and Related Observables}
\label{subsec:dsigmadx}

Next we study the differential cross section $\frac{d\sigma}
{d\cos\theta}(\mu,M_2,\tanb,\msnu)$ of chargino production and consider
associated observables.  As we will emphasize below, only two
combinations of the parameters $(\mu,M_2,\tanb,\msnu)$ can be easily
extracted even theoretically.  One of these is proportional to the
total cross section $\stot$, while the other is proportional to the
production forward-backward asymmetry $\afbchi$.  To a first
approximation, which we will show to be sufficiently accurate, every
other quantity that depends only on $\frac{d\sigma}{d\cos\theta}$ gives
information which is equivalent to that contained in $\stot$ and
$\afbchi$. Unfortunately, with the exception of $\stot$, none of these
quantities is observable, since the direction of the charginos cannot
be fully reconstructed.  The angular distributions of the visible
particles depend on the chargino decay vertices as well as on
$\frac{d\sigma}{d\cos\theta}$, and although they are of great interest,
they are not amenable to analytic study.  They can be investigated by
Monte Carlo simulations, but this is beyond the scope of the present
work.  Our approach therefore only allows us to extract a single
combination of the four underlying parameters that determine the
differential cross section.

We first present the differential cross section for chargino production
and analyze its dependence on the various SUSY parameters in the region
of parameter space given by Eq.~(\ref{bounds}). Formulae for the
differential cross section and total cross section have been given in
many previous studies \cite{srchs,Chia,Btl1,Btl2,Ahn,LeNLC}. Here we
present the differential cross section for completeness and in a form
that allows us to highlight certain properties that have particular
relevance to our analysis.

The cross section is built from the three ingredients in
Fig.~\ref{fig:feynprod}.  The couplings in the virtual photon diagram
are, of course, independent of SUSY parameters. To compute the virtual
$Z$ diagram, we need the couplings of chargino currents to the $Z$. The
virtual $Z$ amplitude for producing chargino states $\chcp_i$ and
$\chcm_j$ is given by

\be\label{Zprod}
\begin{array}{rl}
\bar{v}(e^+)&ig \gamma^{\mu} \frac{\displaystyle 1}{\displaystyle
\cosw} \left[(\frac{1}{2}-\sinsqw ) P_L+(-\sinsqw )P_R\right] u(e^-)
\times \\
&\bar{u}(\chcp_i) ig \gamma_{\mu} \frac{\displaystyle 1}{\displaystyle
\cosw} \left[{\bf O'_L}_{ij}  P_L+{\bf O'_R}_{ij}P_R\right] u(\chcm_j)
\times \frac{\displaystyle -i}{\displaystyle s-\mz^2} \ ,
\end{array}
\ee
where the dependence on the SUSY parameters is through the quantities
that are conventionally labeled ${\bf O'_L}_{ij}$ and ${\bf O'_R}_{ij}$
(see first reference in \cite{Reviews}).  These are the combinations of
{\bf U} and {\bf V} matrices given by

\be\label{oprime}
\begin{array}{rcl}
{\bf O'_L}_{ij} &\equiv& -{\bf V}_{i1} {\bf V}^{\ast}_{j1} -
\frac{1}{2} {\bf V}_{i2} {\bf V}^{\ast}_{j2} + \delta_{ij} \sinsqw\\
{\bf O'_R}_{ij} &\equiv& -{\bf U}^{\ast}_{i1} {\bf U}_{j1} -
\frac{1}{2} {\bf U}^{\ast}_{i2} {\bf U}_{j2} + \delta_{ij} \sinsqw \ ,
\end{array}
\ee
where the indices $i,j$ are 1 for the lighter chargino and 2 for the
heavier. As we will only be concerned with the lighter chargino, we
define $O'_L \equiv {\bf O'_L}_{11}$ and $O'_R \equiv {\bf O'_R}_{11}$.
After a Fierz transformation, the $\snu$ exchange diagram is also of a
similar form, and its amplitude is given by

\be\label{snuprod}
- \frac{1}{2}|{\bf V}_{11}|^2 \, \bar{v}(e^+) ig \gamma^{\mu} P_L
u(e^-) \times \bar{u}(\chcp_i) ig \gamma_{\mu} P_L u(\chcm_j) \times
\frac{\displaystyle -i}{\displaystyle t-\msnu^2} \ .
\ee

Combining these three contributions, one finds that the differential
cross section for chargino production from unpolarized \epem\ beams in
units of R is given by

\be\label{dsig}
\frac{d \sigma}{dx} = \sum_{i=L}^{R} \frac{3v}{32}
\left[ R_i^2 (1-vx)^2 + S_i^2 (1+vx)^2 + 2 R_i S_i (1-v^2)
\right] \ ,
\ee
where the sum is over the two $e^-$ helicities, $v$ is the chargino
velocity, and $x = \cos \theta$, the cosine of the angle between the
positive chargino $\chcp_1$ and the positron beam. The variables
$R_{L,R}$ and $S_{L,R}$ are

\be\label{RS}
\begin{array}{rl}
R_{L}\equiv&1-K_{L} O'_{L} - c_{\snu} \ ,\\
S_{L}\equiv&1-K_{L} O'_{R} \ ,\\
R_{R}\equiv&1-K_{R} O'_{R} \ ,\\
S_{R}\equiv&1-K_{R} O'_{L} \ ,
\end{array}
\ee
where $K_{L,R}$ are constants associated with $Z$ production from
$e^-_{L,R}$,

\be\label{KLR}
\begin{array}{rl}
K_L\equiv& \frac{{\displaystyle s}}{{\displaystyle s-M_Z^2}}
\frac{{\displaystyle 1}}{{\displaystyle \sinsqw \cossqw}}
\left( \frac{1}{2}-\sinsqw \right) \ ,\\
K_R\equiv& \frac{{\displaystyle s}}{{\displaystyle s-M_Z^2}}
\frac{{\displaystyle 1}}{{\displaystyle \sinsqw \cossqw}}
\left( -\sinsqw \right) \ ,
\end{array}
\ee
and the $\snu$ diagram contribution is given by

\be\label{csnu}
c_{\snu}\equiv\frac{2 |{\bf V}_{11}|^2}{\sinsqw [1-2 v x + v^2+4\msnu
^2 / s]} \ .
\ee

We now analyze these formulae in some detail to determine what can be
learned experimentally.   We will show that (a) in most of the allowed
parameter space only two combinations of $O'_L$, $O'_R$, $|{\bf
V}_{11}|^2$ and $\msnu$ contribute significantly to the unpolarized
cross section; (b) the quantities most sensitive to these two
combinations are the total cross section and the forward-backward
production asymmetry; and (c) no other quantities involving the
unpolarized cross section can be found  that add significantly to our
knowledge. We will illustrate these points using a perturbative
expansion of the differential cross section in variables which we will
define below.  While the expansion is not always valid, we have found
numerically that the conclusions that we draw in the perturbative
regime hold throughout the allowed parameter space.

We now identify the small quantities in which to do perturbation
theory. We define

\be\label{smallquants}
\begin{array}{rl}
\bar{K} \equiv\half(K_L+|K_R|)\ ; &\ \kappa\equiv\half(K_L-|K_R|)\ ; \\
\\
\bar{O}\equiv\half(O'_L+O'_R)\ ; &\ \omega\equiv\half(O'_L-O'_R)\ ; \\
\\
\ \xi_0\equiv&\frac{\ds s}{\ds \msnu^2} \ .
\end{array}
\ee
{}From Eq.~(\ref{KLR}) we see that $K_L\approx -K_R$, so
$\bar{K}\gg\kappa$. Specifically, for $\sqrt{s}=190\gev$,
$\bar{K}\approx 1.8\gg\kappa\approx 0.13$, independently of all SUSY
parameters.  From Sec.~\ref{sec:susyparam} and Eq.~(\ref{oprime}) it
follows that $O'_L\approx O'_R$ in both the gaugino and Higgsino
regions, so that in most cases $\bar{O}$, which runs between
$-1+\sinsqw$ in the gaugino region and $-1/2+\sinsqw$ in the Higgsino
region, has absolute value much larger than $\omega$, which is zero in
the gaugino and Higgsino regions and whose absolute value never exceeds
$0.2$. The last small quantity, $\xi_0\equiv{s}/{\msnu^2}$, is small
for sneutrino masses much larger than $200\gev$.  Because of the form
of Eq.~(\ref{csnu}), the expansion can be slightly improved by using
not $\xi_0$ but

\be\label{xidefn}
\xi\equiv \frac{\ds \xi_0}{\ds 1 +(1+v^2)\xi_0/4} \ .
\ee

In the perturbative regime, the unpolarized differential cross section
${d \sigma}/{dx}$ is conveniently written

\be\label{dsigpert}
\frac{d \sigma}{dx} = \frac{3v}{32} \sum_{k=0}^\infty \ x^k A_k \ ,
\ee
where, to first order in $\xi$, $\kappa$ and $\omega$,

\be\label{Aicoeffs}
\begin{array}{rcl}
A_0 & \approx &  4 (2-v^2) \left[1+\bar{K}^2\bar{O}^2- 2\bar{O}\kappa
-\frac{\displaystyle |{\bf V}_{11}|^2}{\displaystyle 4\sinsqw}
(1-\bar{K}\bar{O})\xi\right] \ , \\
\\
A_1 & \approx  & 16v\left[\bar{K}\omega+\frac{\displaystyle |{\bf
V}_{11}|^2} {\displaystyle 8\sinsqw} (1-\bar{K}\bar{O})\xi \right]
\ , \\
\\
A_2 & \approx & \frac{\ds v^2}{\ds 2-v^2}A_0
\ , \\
\\
A_k & \sim &
\order\left(\frac{\displaystyle |{\bf V}_{11}|^2 }
{\displaystyle 4 \sinsqw} v^k \xi^{k-1}\right) \ , k>2
\ .
\end{array}
\ee
(Recall that $|{\bf V}_{11}|^2\leq 1$.) Thus, to a first approximation
it is impossible to measure separately the four quantities of interest
--- $\bar{O}$, $\omega$, $|{\bf V}_{11}|^2$, and $\xi$  --- since the
cross section effectively depends only on $A_0\propto A_2$ and $A_1$.
As can easily be seen from the angular dependence of
Eq.~(\ref{dsigpert}), these coefficients are most sensitively probed by
the total cross section

\be\label{Atostot}
\stot \approx\frac{v}{8}\left(\frac{3-v^2}{2-v^2}\right)A_0
\ee
and the forward-backward asymmetry

\be\label{Atoafb}
\afbchi \approx \frac{3A_1}{4A_0}\left(\frac{2-v^2}{3-v^2}\right) \ .
\ee
The quantities $A_k$ for $k>2$ are greatly suppressed in the
perturbative regime and cannot give additional useful information.

Perhaps surprisingly, the basic conclusions of this perturbative
analysis are correct for the entire accessible region of parameter
space.   The coefficients $A_k$, $k>2$, become substantial and the
relation $A_0\propto A_2$ breaks down only for $|{\bf V}_{11}|$ near
one and for large $\xi$, that is, in the gaugino region with a
sneutrino mass near $100\gev$. However, even in this case we find
numerically that it is extremely difficult to extract additional
information from the differential cross section.  For example, consider
a simple variable which is orthogonal to $\afbchi$ and linearly
independent of $\stot$:

\be\label{Acentral}
A_{{\rm central}}^{\chc} =
\frac{\ds \int_{-1/2}^{1/2} dx \ \frac{d \sigma}{dx}}{\ds \stot} \ .
\ee
In the perturbative regime this variable is simply

\be\label{Acentralpert}
A_{{\rm central}}^{\chc} \approx A_{{\rm central}}^{(0)\chc}
\equiv\frac{1}{16}\frac{24-11v^2}{3-v^2} \ .
\ee
which follows from Eqs.~(\ref{dsigpert}) and (\ref{Aicoeffs}).
Deviations from this prediction would provide new information beyond
$\stot$ and $\afbchi$.  In Fig.~\ref{fig:Ac} the ratio $A_{{\rm
central}}^{\chc} / A_{{\rm central}}^{(0)\chc}$ is shown for two light
sneutrino masses $\msnu = 200 \gev$ and 100 GeV. The figure indicates
that $A_{{\rm central}}^{\chc}$ is within a few percent of the
perturbative prediction everywhere except in the gaugino region for
extremely small sneutrino mass and large chargino mass, where a ten
percent deviation is seen. In principle this small effect would be
useful, but $A_{{\rm central}}^{\chc}$, like $\afbchi$, is not directly
observable and must be estimated through its correlation with some
observable quantity.  Even were this somehow to be overcome through
Monte Carlo simulations, the chargino cross section at LEP II provides
us with at most a few thousand events; statistical errors on the
measured $\stot$ and $\afbchi$ are already several percent. We
therefore cannot expect to gain much from this variable.

We have searched for other possible observables, but have found none
with both large variation and small experimental uncertainties; the
behavior of $A_{{\rm central}}^{\chc}$ is typical.  Statistical errors
alone make any of these variables difficult to use; but the
impossibility of directly measuring the chargino momentum axis greatly
complicates the determination of any variable based on distributions in
chargino production angle.  For our purposes, then, the total cross
section and the forward-backward chargino production asymmetry are the
only potentially useful quantities stemming from the unpolarized
differential cross section. In the next subsections we discuss these
two quantities in detail.

\subsection{Total Cross Section}
\label{subsec:sigmatotal}

In this subsection we analyze the total cross section in detail. We
will find that, as has been noted previously in the literature
\cite{srchs,Chia,Btl1,Btl2,Aachen,Chen,Grvz2,Ahn,Grvz,JLC,Fujii,LeLEP},
charginos are produced in the thousands in most areas of the accessible
parameter space, and the production cross section is highly sensitive
to sneutrino mass.

{}From Eqs.~(\ref{Aicoeffs}) and (\ref{Atostot}) it can be seen that in
the gaugino region, for small $\xi$, the cross section is approximately

\be\label{stotG}
\begin{array}{rl}
\stot\approx&
f(v) \left\{ 1 +
     \left[   \frac{\ds s}{\ds s-\mz^2}\right]^2
     \left(\frac{\ds 1}{\ds 4\sinsqw}\right)^2
     - \frac{\ds |V_{11}|^2}{\ds 4\sinsqw}
     \left[ 1+\frac{\ds s}{\ds s-\mz^2}
     \frac{\ds 1}{\ds 4\sinsqw}\right]\xi\right\}\\
\\
\approx& ( 3.2 - 2.8 \xi)f(v)\ {\rm R}\
\approx  (8.8 \picob)\ (1-0.9\xi)\ f(v) \ ,
\end{array}
\ee
while in the Higgsino region it is
\be\label{stotH}
\begin{array}{rl}
\stot\approx&
f(v)\left\{ 1 +
      \left[   \frac{\ds s}{\ds s-\mz^2}\right]^2
        \left(\frac{\frac{1}{2}\ds -\sinsqw}{\ds
        1-\sinsqw}\right)^2\right\}\\
\\
\approx&
    1.3\ f(v)\ {\rm R} \  \approx (3.6 \picob)\ f(v) \ ,
\end{array}
\ee
where $f(v) = \half v (3-v^2)$ rises from zero at threshold to one at
high energy, $\xi$ is defined in Eq.~(\ref{xidefn}), and where we have
taken $\sqrt{s}=190\gev$, for which one unit of R is approximately
$2.75\picob$. Strong sensitivity to $\msnu$ is found in the gaugino
region \cite{Btl1,Btl2,Ahn}, but disappears altogether in the Higgsino
region. In the large $\msnu$ limit both expressions are entirely
determined by gauge invariance, but the event rate is two and a half
times higher in the gaugino region. Notice that while a large cross
section is a signal of the gaugino region, a small one can occur both
in the Higgsino region and, if the sneutrino is light, in the gaugino
region.

These features can all be seen in Fig.~\ref{fig:sigma}, where chargino
production cross sections for $\msnu= 1 \tev$ and 150 GeV are plotted
in picobarns. Because the cross section plots do not change
substantially for different $\tanb$, the plots are presented for
$\tanb=4$ only. We see that, in a sample of $1 \ifb$, LEP II will
produce thousands of chargino events in most of the accessible regions
of parameter space. Contrasting Figs.~\ref{fig:sigma}a and b, one can
also see the strong dependence on $\msnu$ noted above. This is shown
more explicitly in Fig.~\ref{fig:sigmasnu}, where we plot the total
cross section at four representative $\mum$ points as a function of
$\msnu$. Clearly the $\snu$ diagram can give a large and destructive
contribution to the cross section. For this reason, the infinite
sneutrino mass limit is neither representative nor conservative and can
lead to a substantial overestimate of the event rate in the gaugino
region.  This in turn could result in overly optimistic claims
concerning the statistical accuracy with which SUSY parameters can be
determined.

Having analyzed the dependence of $\stot$ on the fundamental SUSY
parameters, we now turn to the issue of how $\stot$ may be extracted
from experiment. To measure $\stot$, it will be necessary to measure
the partial cross sections of chargino production in at least two of
the hadronic, mixed, and leptonic decay modes. The partial cross
sections are given by

\be\label{partsig}
\begin{array}{rl}
\sigma_{\rm leptonic} &= B_l^2 \stot \\
\sigma_{\rm mixed} &= 2 B_l (1- B_l) \stot \\
\sigma_{\rm hadronic} &= (1-B_l)^2 \stot \ ,
\end{array}
\ee
where $B_l$ is the leptonic branching fraction defined in
Eq.~(\ref{fourobs}).  As these three partial cross sections depend on
only two variables, they must satisfy the constraint

\be\label{partsig2}
\sigma_{\rm leptonic}\sigma_{\rm hadronic} =
\fourth\sigma_{\rm mixed}^2 \ .
\ee
This relation is not dependent on the details of the chargino decay
process; if it is not satisfied, it indicates a problem with the
estimated detection efficiencies in one or more of the modes.  (Such a
discrepancy could stem  either from an experimental problem  or from
physics beyond our minimal assumptions that is not included in the
Monte Carlo simulation---for example, signal from an additional and
unexpected supersymmetric particle.)  To determine $\stot$ it is best
to use all three partial cross sections, subject to the constraint in
Eq.~\ref{partsig2}, but we will only use the two with the smallest
errors ($\sigma_{\rm mixed}$ and one of the other two). We note that in
measuring $\stot$, we obtain also $B_l$, and it is therefore natural to
examine this observable, as we will do in Sec.~\ref{subsec:BrLepton}.

\subsection{Chargino Forward-Backward Asymmetry}
\label{subsec:AFB}

The forward-backward asymmetry $\afbchi$ of chargino production is
theoretically attractive, since it can be computed analytically and
depends on the four parameters $(\mu,M_2,\tanb,\msnu)$ in a way which
is quite different from the total cross section. In the large $\msnu$
limit, $\afbchi$ is negative in the mixed region and negligible
elsewhere, as can be seen from Eqs.~(\ref{Aicoeffs}) and
(\ref{Atoafb}); however, a light sneutrino, which appears in a
$t$-channel diagram, can give a large positive contribution to
$\afbchi$ in the gaugino and mixed regions.  This effect is seen in
Fig.~\ref{fig:afbfig}.

Unfortunately, $\afbchi$ cannot be directly measured; the velocities of
the two charginos cannot be reconstructed because the two LSPs are
invisible.  Let us consider what might be possible to observe in these
events.  Since the forward-backward asymmetry is odd under charge
conjugation, we must discover which chargino is positively charged. The
sign of the charge of a single chargino can be determined in a leptonic
decay from the charge of the lepton, but is more difficult to measure
in a hadronic decay. We must also determine the momentum axis of the
chargino. The visible particles in the decay of a chargino can in
certain cases indicate the chargino momentum. For example, if the
chargino is moving at relativistic velocities, the decay products are
highly boosted along the chargino momentum axis.  Alternatively, if in
the chargino rest frame the total visible momentum, averaged over many
events, is distributed isotropically, then the average visible momentum
in any frame will give the chargino momentum direction.  Unfortunately,
neither of these cases applies here; the velocity of the chargino is
generally semi- or non-relativistic, and the decays are often far from
isotropic.

Still, we might hope that the distribution of the dijet momenta in the
hadronic decays of charginos would give a reasonable estimate of the
chargino momentum distribution and  could be used to measure $\afbchi$.
Specifically, in mixed mode chargino decays, we may use the dijet
momentum to estimate the chargino momentum axis, and the charge of the
lepton to determine the direction of the positively charged chargino.
The dijet forward-backward asymmetry $\afbjj$ found in this way might
well be correlated with $\afbchi$.  Unfortunately, the correlation is
often very weak. As described in Sec.~\ref{sec:studies}, for each case
study, we have explored via Monte Carlo simulation the region of SUSY
parameter space which gives the observed $\mchargino$, $\mLSP$, $\stot$
and $B_l$. We have found that for these points in parameter space it is
impossible to estimate $\afbchi$ using $\afbjj$ without additional
information about the decay process. In fact, the variation of $\afbjj$
around $\afbchi$ is so large that it indicates that $\afbjj$ has strong
dependence on the parameters in the decay vertices, and possibly can be
used to determine them. However, as analytic formulae for this variable
are unavailable, and since experimental cuts must be included, this
will require a detailed Monte Carlo simulation covering all of the
allowed parameter space, which we do not attempt here.  (We note that
analytic formulae may be found near the threshold of chargino
production, which may permit the separation of $\afbchi$ from the decay
vertices \cite{decaystudy}.)

Additionally, we have considered a range of cuts on the data to try to
improve the correlation between $\afbjj$ and $\afbchi$. In particular,
the angle between the dijet and chargino momenta may be determined
simply from momentum conservation.  By cutting away events where this
angle is large, we obtain a sample of events where the dijet and
chargino momenta point in roughly the same direction. In some cases the
correlation is slightly improved for this sample, but this is
counterbalanced by the loss of statistics. We note that this sample of
events might be useful for other purposes, such as determining the
parameters in the decay vertices.

We find, then, that it is impossible to measure $\afbchi$ directly.
However, the observable $\afbjj$, although too complicated to explore
without a Monte Carlo simulation, contains a mixture of information
about $\afbchi$ and the decay vertices that will certainly be of
interest.  We will discuss this further in Sec.~\ref{sec:studies}.

\subsection{Polarization Asymmetry}
\label{subsec:ALR}

While the unpolarized cross section essentially contains only two
separable parameters, a third independent observable,

\be\label{RLratio}
\frac{\ds\sigma_R}{\ds \sigma_L} \equiv
\frac{\ds \sigma(e^-_R e^+_L \rarr \chcp\chcm)}
{\ds \sigma(e^-_L e^+_R \rarr \chcp\chcm)} \ ,
\ee
can be studied when polarized electron beams are available.
Unfortunately, this is not expected to be the case at LEP II.  Still,
this observable has important implications for chargino threshold
studies.

The ratio ${\sigma_R}/{\sigma_L}$ is less than $15\%$ throughout the
allowed region of parameter space, and so, even with unpolarized beams,
charginos are always produced largely by left-handed electrons
\cite{Chia,LeNLC,LeLEP}. In the large $\msnu$ limit,
${\sigma_R}/{\sigma_L}\sim 1\%$ in the gaugino and $\sim 15\%$ in the
Higgsino region; it varies widely in the mixed region, but is generally
small.  A light sneutrino can increase the ratio in the gaugino and
mixed region.  In Fig.~\ref{fig:sigmalr} we show
${\sigma_R}/{\sigma_L}$ for $\msnu=150\gev$, for which the ratio takes
values near its maximum. Nevertheless, we see that
${\sigma_R}/{\sigma_L}$ is still small, and is approximately $2\%$ in
much of the gaugino region. While this ratio is independent of $\stot$
and $\afbchi$, it is not uncorrelated with the total cross section
$\stot = {\sigma_R}+{\sigma_L}$, since $\msnu$ enters only in
${\sigma_L}$ and ${\sigma_L}\gg{\sigma_R}$.  Still, the correlation is
imperfect and it is possible to gain some amount of new information.

It is possible to exploit experimentally the theoretical prediction
that ${\sigma_R}/{\sigma_L}$ is small.  Near threshold, this implies
that charginos are produced dominantly polarized along the beam axis,
and the study of their decays is therefore greatly simplified.  As the
combinations of the SUSY parameters that enter the decay are different
from those that enter the production process, measurements at threshold
can give valuable information to supplement the analysis presented in
this paper \cite{decaystudy}.  In this study we assume a fixed
center-of-mass energy, which means that in much of the parameter region
the charginos have substantial velocities, and near-threshold analysis
is not applicable.

\subsection{Leptonic Branching Fraction}
\label{subsec:BrLepton}

To measure the total cross section, we must determine at least two of
the partial cross sections for the hadronic, mixed, and leptonic modes.
These also provide us with a measurement of the leptonic branching
fraction $B_l(\mu,M_2,\tanb, M_1,\mslep,\msq )$, which can vary greatly
in the SUSY parameter space \cite{branchingratios}.  We now analyze the
dependence of $B_l$ on the underlying SUSY parameters.

As discussed earlier, charginos decay to a neutralino and either two
hadrons or two leptons.  The hadronic decays occur via intermediate $W$
bosons and squarks (Fig.~\ref{fig:feyndecay}). The decay $\chcp_1 \rarr
\chn_1 u \bar{d}$ through a virtual $W$  has amplitude

\be\label{decaytoW}
\bar{u}(\chn_1)ig\gamma^\mu(O_L P_L+O_R P_R)u(\chcp_1) \times
\bar{u}(u)\frac{-ig}{\sqrt{2}}\gamma_\mu P_L v(\bar{d}) \times
\frac{-i}{[(p_u+p_{\bar{d}})^2-\mw^2]} \ ,
\ee
where

\be\label{OLORdefn}
\begin{array}{rcl}
O_L &\equiv& {\bf N}_{12} {\bf V}^*_{11}-\frac{1}{\sqrt{2}} {\bf
N}_{14} {\bf V}^*_{12} \\
O_R &\equiv& {\bf N}^*_{12} {\bf U}_{11}+\frac{1}{\sqrt{2}} {\bf
N}^*_{13} {\bf U}_{12} \ .
\end{array}
\ee
The decay mediated by a left-handed up squark has amplitude

\be\label{decaytouL}
\begin{array}{rl}
\bar{v}(\chcp_1)(-ig)&{\bf V}^*_{11}  P_L v(\bar{d}) \\
&\times
\bar{u}(u) (-ig\sqrt{2}) (Y_Q \tanw {\bf N}_{11}+ I_u
{\bf N}_{12})P_R v(\chn_1)
\frac{\ds i}{\ds [(p_{\chn}+p_{u})^2-m_{\tilde{u}_L}^2]} \ ,
\end{array}
\ee
while that mediated by a left-handed down squark takes the form

\be\label{decaytodL}
\begin{array}{rl}
- \bar{u}(u) (-ig)&{\bf U}_{11}  P_R u(\chcp_1) \\
&\times
\bar{u}(\chn_1)(-ig\sqrt{2}) (Y_Q \tanw {\bf N}^*_{11}
+ I_d {\bf N}^*_{12}) P_L v(\bar{d})
\times \frac{\ds i}
{\ds [(p_{\chn}+p_{\bar{d}})^2-m_{\tilde{d}_L}^2]} \ .
\end{array}
\ee
The overall minus sign of the down squark amplitude results from the
odd permutation of the spinors of the down squark amplitude relative to
the spinors of the other two diagrams. In these formulae
$Y_Q=\frac{1}{6}$ is the hypercharge of left-handed quarks, while
$I_u(I_d)=\half(-\half)$ is the weak isospin of up (down) quarks. We
remind the reader that only the first two generations of squarks
participate in chargino decays, since we have assumed that squark
mixing angles are small and that the chargino is lighter than the top
quark.

In Eqs.~(\ref{decaytouL}) and (\ref{decaytodL}) we have omitted the
couplings of squarks to the Higgsino component of the gauginos, because
they are suppressed by $m_d/(\mw\cos\beta)$ or $m_u/ (\mw\sin\beta)$,
and are therefore negligible. As discussed in Sec.~\ref{sec:susyparam},
the right-handed squark diagrams are similarly suppressed and may also
be ignored.  (We note, however, that the very small effects of the
Higgsino couplings and the right-handed sfermion diagrams are included
in our Monte Carlo simulation.)

The squark diagram contributions may be Fierz transformed into the same
form as the $W$ diagram amplitude; the full hadronic decay amplitude
may then be written as

\be\label{decaytoall}
\frac{i^3g^2}{\sqrt{2}} \
\bar{u}(\chn_1)\gamma^\mu(D_L P_L+D_R P_R)u(\chcp_1) \times
\bar{u}(u)\gamma_\mu P_L v(\bar{d}) \ ,
\ee
where

\be\label{DLdefnq}
D_L(q)=\frac{O_L}{(p_u+p_{\bar{d}})^2-\mw^2}+
\frac{{\bf V}^*_{11}(\frac{1}{6} \tanw {\bf N}_{11}+\half {\bf
N}_{12})} {(p_{\chn_1}+p_{u})^2-m_{\tilde{u}_L}^2} \ ,
\ee
and

\be\label{DRdefnq}
D_R(q)=\frac{O_R}{(p_u+p_{\bar{d}})^2-\mw^2}- \frac{{\bf
U}_{11}(\frac{1}{6} \tanw {\bf N}^*_{11}-\half {\bf N}^*_{12})}
{(p_{\chn_1}+p_{\bar{d}})^2-m_{\tilde{d}_L}^2} \ .
\ee
Notice that up(down)-type squarks contribute only to $D_L(D_R)$.

For leptonic decays, in which all three generations of sleptons
participate, the full amplitude has the same form as
Eq.~(\ref{decaytoall}) but with hypercharge equal to $-\half$:

\be\label{DLdefnl}
D_L(l)=\frac{O_L}{(p_\nu+p_{\bar{e}})^2-\mw^2}+
\frac{{\bf V}^*_{11}(-\half \tanw {\bf N}_{11}+\half {\bf N}_{12})}
{(p_{\chn_1}+p_{\nu})^2-m_{\tilde{\nu}_L}^2} \ ,
\ee
and

\be\label{DRdefnl}
D_R(l)=\frac{O_R}{(p_\nu+p_{\bar{e}})^2-\mw^2}-
\frac{{\bf U}_{11}(-\half \tanw {\bf N}^*_{11}-\half {\bf N}^*_{12})}
{(p_{\chn_1}+p_{\bar{e}})^2-m_{\tilde{e}_L}^2} \ .
\ee
As above, the isospin $+\half$ ($-\half$) left-handed slepton
contributes to $D_L(D_R)$.

In the approximation that the momentum dependences of the $W$ and
scalar propagators are ignored, the partial widths of the chargino can
be written down in closed form.  In this limit, the width, in terms of
$r=\mLSP /\mchargino$, is

\be\label{DecayWidthA}
\Gamma(\chcp\rarr \chn f\bar{f}) \approx
N_f N_c\frac{\ds g^4 m_{\chc_1}^5}{\ds 3072 \pi^3} G(r)
\Big[D_L(f)^2+D_R(f)^2 - g(r)D_L(f) D_R(f)\Big] \ ,
\ee
where

\be\label{DecayWidthA2}
\begin{array}{rl}
G(r)=&1-8r^2-24r^4 \log r +8r^6-r^8 \ ,\\
g(r)=&4r (1+9r^2+12r^2\log r - 9 r^4 + 12r^4 \log(r) - r^6)/G(r) \ ,
\end{array}
\ee
and where for hadrons (leptons), $f=q$ ($f=l$), the number of flavors
$N_f$ is $2$ ($3$), and the number of colors $N_c$ is $3$ ($1$). The
function $g(r)$ is well approximated by $1-(1-r)^4$, while for $r\rarr
1$, $G(r)\sim(1-r)^5$. The ratio of leptonic to hadronic  branching
fractions is given by

\be\label{Rlhratio}
\frac{B_l}{1-B_l} =
   \half\frac{D_L^2(l)+D_R^2(l)-g(r)D_L(l) D_R(l)}
        {D_L^2(q)+D_R^2(q)-g(r)D_L(q) D_R(q)} \ .
\ee

The dependence of $B_l$ on the parameters of supersymmetry is quite
complicated, and there are few regions of parameter space in which a
useful perturbative expansion may be performed.  However, it is
possible to make some broad statements about its behavior. In
Fig.~\ref{fig:Bl} we present $B_l$ for three different values of the
parameters. Note that $B_l$ is unlike the other three variables we have
looked at, in that it can have strong $\tanb$ dependence in the gaugino
region.  In the following we will discuss some of the most notable
features in the figures.

In the Higgsino region of Figs.~\ref{fig:Bl}a-c, the branching fraction
is approximately $\frac{1}{3}$.  This is a general phenomenon. As the
Higgsino region is approached from the gaugino region, the couplings of
the chargino to squarks and sleptons decrease, while the couplings to
the $W$ boson increase. In the far Higgsino region, the sfermion
couplings to the chargino are completely negligible, so chargino decay
is dominated by virtual $W$ bosons for which the branching ratio $B_l$
is $\frac{1}{3}$.  (Recall that we include in $B_l$ {\it all} chargino
decays to $\tau$ leptons, even if the $\tau$ itself decays
hadronically.) Even outside the Higgsino region, effects of heavy
squarks and sleptons are suppressed relative to the intermediate $W$
boson; the large sfermion mass limit again leads to
$B_l\approx\frac{1}{3}$.  Thus,  we cannot draw any conclusions if
$B_l\approx\frac{1}{3}$, but if $B_l$ is not equal to $\frac{1}{3}$, it
immediately rules out both the Higgsino region and ultraheavy sfermions.

In the gaugino region, where $|\mu|\gg M_2$, the matrix elements ${\bf
U}_{11}, {\bf V}_{11}$ and ${\bf N}_{11}$ are all very close to unity,
while the couplings $O_L$ and $O_R$ of the chargino to the $W$ are of
order the mixing angle ${\bf N}_{12}$, as are the isospin-dependent
terms in $D_{L,R}(l,q)$.  All other relevant elements of the ${\bf U}$,
${\bf V}$ and ${\bf N}$ matrices are small. A perturbative
diagonalization of the matrix in Eq.~(\ref{neumass}) shows that in the
range of parameters we consider,

\be\label{gauginoWchnmix}
{\bf N}_{12} \sim \Big(\frac{M_Z}{\mu}\Big)^n \ ,
\ee
where $n=1$ for $M_Z\tanb\ll|\mu|$, and $n=2$ for $M_Z\tanb\gg|\mu|$.
The latter reflects the $\mu\rarr -\mu$ symmetry for large $\tanb$.

It follows that for very large $|\mu|$ and sufficiently small slepton
and squark masses, the $W$ diagram and the isospin terms in the squark
and the slepton diagrams are negligible, as can be seen from
Eqs.~(\ref{decaytouL}) and (\ref{decaytodL}).  Keeping only the
hypercharge terms in $D_{L,R}(q,l)$ and noting $(Y_l/Y_q)^2=9$, one
finds

\be\label{BlfarG}
B_l\approx\frac{\ds \msq^4}{{\ds \msq^4}+\frac{2}{9}{\ds \mslep^4}}
\ee
in the far gaugino region.  This is larger than $\frac{1}{3}$ for each
of the cases plotted in Fig.~\ref{fig:Bl}, in which the growth of $B_l$
at large $|\mu|$ is evident.  Notice, however, that the growth is much
faster in the large $\tanb$ case (Fig.~\ref{fig:Bl}c) than in the small
$\tanb$ case (Figs.~\ref{fig:Bl}a,b).  Note also that the figures at
$\tanb=2$ with two different squark masses (Figs.~\ref{fig:Bl}a,b) have
$B_l$ contours that are shifted by a constant factor.  We explain these
features below.

It is possible to estimate roughly where the crossover occurs from
isospin dominated chargino decays, which are close to $B_l=\frac{1}{3}$
for $\msq\sim\mslep$, to hypercharge dominated decays, which have a
$B_l$ given by Eq.~(\ref{BlfarG}).  We first consider small $\tanb$.
Take $|\mu|\gg M_Z\tanb$, but in the near gaugino region, where
$B_l\sim\frac{1}{3}$. Now raise $|\mu|$ until ${\bf N}_{12}$ is so
small that the hypercharge terms in Eqs.~(\ref{decaytouL}) and
(\ref{decaytodL}) dominate the decay amplitudes. Since, for small
$\tanb$, ${\bf N}_{12}\sim M_Z/\mu$, we may estimate that this occurs
for

\be\label{bllowtanb}
|\mu|\sim \frac{1}{\tanw |Y_f|}\frac{m_{\tilde{f}}^2}{\mw^2}M_Z \ .
\ee
This rough estimate is accurate only within a factor of four or so.
Since $\tanw |Y_l|$ is approximately $1/4$ for leptons and $1/12$ for
quarks, the required $|\mu|$ is generally larger than a TeV except for
the lightest sfermions, and is smaller for sleptons than for squarks of
the same mass.  We therefore expect that for small $\tanb$, $B_l$ will
not stray too far from $\frac{1}{3}$. In the limit $\tanb=1$ we may
quantify this; for large squark and slepton masses, the leading
hypercharge terms in the $D_{L,R}(q,l)$ cancel, with the isospin terms
giving

\be\label{lhratioisospin}
{B_l} \approx \frac{1}{3}\Bigg[1+
\frac{2}{3}\left(\frac{\mw^2}{\mslep^2}-\frac{\mw^2}{\msq^2}\right)
\Bigg] \ ,
\ee
to first order in $\mw^2 /\mslep^2$ and $\mw^2 /\msq^2$. This shows
that even within the isospin terms there is sensitivity, independent of
$\sinw$ and $\tanb$, to the difference between $\msq$ and $\mslep$.
This is a general feature of small $\tanb$, as is reflected in
Figs.~\ref{fig:Bl}a and b, where the $B_l$ contours differ by
approximately 5\% and are consistent with Eq.~(\ref{lhratioisospin}).

Suppose instead that $\tanb$ is large and take $|\mu|\ll M_Z\tanb $,
again in the near gaugino region with $B_l\sim\frac{1}{3}$. Let us
again consider increasing $|\mu|$.  Now ${\bf N}_{12}\sim (M_Z/\mu)^2$
vanishes much more quickly as $|\mu|$ grows; the crossover to
Eq.~(\ref{BlfarG}) occurs near

\be\label{blhightanb}
|\mu|\sim \frac{1}{\sqrt{\tanw |Y_f|}}\frac{m_{\tilde{f}}}{\mw}M_Z
\ee
The dependence on the sfermion mass is now linear, and the coefficient
has become smaller as well.  For large $\tanb$, the leptonic branching
fraction will therefore deviate strongly from $\frac{1}{3}$ at much
smaller values of $|\mu|$ than is the case for small $\tanb$.  This is
clearly seen in the difference between Figs.~\ref{fig:Bl}a and c.  We
note that in the near gaugino region, $B_l$ is also shifted by the
correction term in Eq.~(\ref{lhratioisospin}) even for large $\tanb$.

Another feature worth noting is a ``pocket'' in which $B_l$ drops below
$\frac{1}{3}$, which occurs in the near gaugino region for negative
$\mu$, where simple perturbation theory is not applicable. In this
region, the slepton diagrams interfere destructively with the $W$
diagram in the decay amplitude. The ``pocket'' can be deep for
intermediate values of $\tanb$, where $B_l$ can take values as low as
10\%. This feature has important implications for the case studies of
Sec.~\ref{sec:studies}.

As noted above, $B_l$ can be determined by measuring the partial cross
sections.  The leptonic branching fraction is given by

\be\label{blsig}
B_l = \frac{2\sigma_{\rm leptonic}}{2\sigma_{\rm leptonic} +
\sigma_{\rm mixed}}
 = \frac{\sigma_{\rm mixed}}{\sigma_{\rm mixed} +
2\sigma_{\rm hadronic}}.
\ee
Again, all three partial cross sections should be employed, though we
will only use the two that are likely to have the smallest errors for
our measurement of $B_l$.  Strictly speaking, the partial cross
sections must be adjusted to account for the fact that $\tau$ leptons
can decay hadronically. As the corrections depend solely on the
well-measured $\tau$ branching fractions, the required adjustments are
very reliable.

\subsection{Other Observables}
\label{subsec:otherobs}

There are many other quantities that could be extracted from the data
which depend on the details of the chargino decay vertices. For $2j+l$
events, distributions of $\theta_l$, the angle between the lepton and
the beam axis, $E_l$, the energy of the lepton, and corresponding dijet
variables $\theta_{jj}$ and $E_{jj}$, are observables that are likely
to yield useful information. Correlations between $l$ and $\bar{l}$
angles in $2l$ events and between lepton and dijet angles in $2j+l$ may
also provide information \cite{Hagiwara}.  However, to obtain analytic
expressions for these variables one would have to convolve the angular
velocity and spin distribution of the production process with the
three-body spin-dependent phase space of the decay. The complexity of
dealing with these observables is regrettable, since the angular
distributions of the chargino decay products are sensitive to
$D_L(q)/D_R(q)$ and $D_L(l)/D_R(l)$ (see Sec.~\ref{subsec:BrLepton}),
which cannot be probed using the quantities $\mchargino$, $\mLSP$,
$\stot$ and $B_l$. Furthermore, the angular distributions might help
determine the angle and spin dependence of the differential cross
section, and might be noticeably affected by the propagators of very
light sfermions.  Restrictions on SUSY parameter space using these
observables will probably require detailed  Monte Carlo simulations,
which we will not attempt to carry out.  It seems plausible to us that
a global maximum likelihood fit to the data, on the basis of a
sufficiently thorough Monte Carlo search, should be able to pick out
information that we have not been able to extract in our analysis; we
will bolster this claim in Sec.~\ref{sec:studies}. It remains to be
seen whether the additional knowledge will lead to significantly
improved constraints on the underlying parameters.

There are other variables worthy of study which allow the assumptions
of our analysis to be tested.  If the LSP is a sneutrino instead of a
neutralino, chargino decays are qualitatively different, since the
two-body decay $\chargino\rarr l\snu$ becomes available. If there is
substantial mixing between any of the right- and left-handed sfermions,
the chargino decay amplitude will no longer have the form of
Eq.~(\ref{decaytoall}); the resulting angular distribution of the
observed fermions will be identifiably different. Mixing of the third
generation of squarks into the other generations could cause many $b$
quarks to appear in the dijets; under our assumptions, very few are
expected.  A significant breakdown of the universal slepton mass
assumption would affect lepton universality in the decays; similarly, a
violation of our universal squark mass assumption for the first two
generations might be detected by studying the abundance of charm quarks
in the dijets.  Lastly, a significant contribution by intermediate
charged Higgs bosons to the chargino decay amplitude would both affect
the angular distributions of the observed fermions and lead to extra
heavy fermions among the decay products.

\subsection{Summary of Observables}
\label{subsec:obssummary}

We have now concluded the discussion of the four primary observables
that will be used in the case studies below.  These observables, with
their dependences on the underlying SUSY parameters, are the chargino
mass $\mchargino (\mu, M_2, \tanb)$, the LSP mass $\mLSP (\mu, M_2,
\tanb, M_1)$, the total cross section $\stot (\mu, M_2, \tanb,
\mslep=\msnu)$, and the leptonic branching fraction $B_l (\mu, M_2,
\tanb, M_1, \mslep, \msq)$.  We have also studied the forward-backward
asymmetry  $\afbchi(\mu, M_2, \tanb, \mslep=\msnu)$ and have explained
why it is difficult to use.

To close this section, we review the sensitivity of these parameters in
the different regions. In the gaugino region, where $|\mu|\gg
M_2,|M_1|$, we are sensitive to all six parameters: $\mchargino\sim
M_2$; $\mLSP\sim |M_1|$; $\stot \approx f(v)\times [3.2-2.8(s/\msnu^2)]
\, {\rm R}$ to leading order in $s/\msnu^2$, where $f(v)$ is a definite
function of the chargino velocity; and $B_l$ is a non-trivial function
which is sensitive to several combinations of SUSY parameters. When
$\mslep$ $(= \msnu)$ and $\msq$ are both very large, then $\stot =
f(v)\times 3.2 \, {\rm R}$ and $B_l=\frac{1}{3}$. In the Higgsino
region, where $|\mu|\ll M_2$, the following relationships hold:
$\mchargino\sim |\mu|$; $\mLSP\sim \min\{|\mu|,|M_1|\}$; $\stot =
f(v)\times 1.3 \,{\rm R}$; and $B_l=\frac{1}{3}$. There is no
sensitivity in this region to $\mslep$ or $\msq$, and very little to
$M_1/M_2$, unless $|M_1|$ is less than or of order $|\mu|$. In the
mixed region, the observables are all complicated functions of the SUSY
parameters, and there are few general statements to be made.

{}From this discussion it can be guessed that we can put the fewest
constraints on parameter space if the physics lies in the Higgsino
region, whereas the gaugino region is more promising for our analysis.
Intermediate results are found in the mixed region. We will see this
explicitly in the case studies in Sec.~\ref{sec:studies}.

\section{Event Simulation and Backgrounds}
\label{sec:simulation}

For this study, chargino events are generated by a simple parton level
Monte Carlo event generator.  For a given set of parameters, we first
calculate the chargino decay width and branching ratios.  The SAGE
subroutines \cite{SAGE} are used to generate three-body final state
momenta and phase space weightings, and the matrix elements for the
decay are calculated with the explicit helicity spinor method, using
subroutines that are patterned after the HELAS package \cite{HELAS}. We
find that typically the chargino width $\Gamma_{\chargino}$ is roughly
1--100 keV. Using the same subroutines, we therefore generate the
six-body events $\epem \rightarrow \chcp_1 \chcm_1 \rightarrow (\LSP
q'q, \LSP l\nu) (\LSP q'q, \LSP l\nu)$ in the zero width approximation
for the intermediate charginos.  In this approximation the total
amplitude factorizes into production and decay amplitudes and is given
by

\be\label{mtot}
\mtot = \sum _{h^+,h^- = -1} ^1 \mprod_{h^+h^-} \mdecay_{h^+}
\mdecay_{h^-} \frac{\pi}{\mchargino \Gamma_{\chargino}},
\ee
where

\be\label{mpieces}
\begin{array}{rl}
\mprod_{h^+h^-} &= \mprod (\epem \rightarrow
\chcp_{h^+}\chcm_{h^-}), \\
\mdecay_{h^{\pm}} &= \mdecay(\chc_{h^{\pm}}
\rightarrow \LSP q'q, \LSP l\nu),
\end{array}
\ee
and $h^{\pm}$ is the helicity of $\chargino$. Without factorization the
amplitude consists of up to 108 diagrams, since 3 diagrams contribute
to the production process and 6 (5) diagrams contribute to each
hadronic (leptonic) decay. (Of course, with the assumptions of
Sec.~\ref{sec:susyparam}, only 3 (3) diagrams contribute substantially
to the hadronic (leptonic) decay.) Factorization allows us to calculate
$\mprod_{h^+h^-}$ and $\mdecay_{h^{\pm}}$ separately, which is a great
simplification and considerably improves the efficiency of the event
generator.  After calculating these amplitudes, we sum over internal
chargino helicities to get $\mtot$, which is then squared and summed
over external helicities to obtain the total differential cross
section. By summing over internal helicities before squaring, we retain
the important spin correlations between production and decay.

In this study, we ignore effects due to the Majorana nature of the
$\LSP$.  Because the $\LSP$ is a Majorana fermion, one should in
principle include for every Feynman diagram a diagram with the $\LSP$
momenta interchanged.  In our Monte Carlo simulation, the momenta are
preferentially picked such that the charginos are very nearly on-shell.
The chargino rest frames are boosted with respect to each other, and
most events are produced with the two LSPs having different momenta.
Thus, in almost all cases, the diagram with interchanged LSP momenta
has chargino momenta that are far out of their narrow Breit-Wigner
peaks and can be ignored.

The event generator was checked in a number of ways.  In the explicit
spinor method, Lorentz invariance is not manifest.  We have checked
that the total amplitude squared is invariant under arbitrary Lorentz
boosts, and this provides a powerful check. In addition, the amplitudes
must transform into their complex conjugates, up to a sign, when all
helicities are reversed, and this was verified as well. The production
cross section was found to agree with the analytic results presented in
Sec.~\ref{sec:observables} and with those previously published
\cite{Btl2} for many sets of parameters. The decay amplitudes were also
verified in a number of ways.  The differential decay width $d\Gamma /
dE d\cos\theta$ was found to agree with muon decay in the appropriate
limit, and also with the decay of a ``muon'' with V+A coupling. This
provides a check of the $W$ diagram and the overall normalization.
Finally, the total decay amplitude, summed over all diagrams, was
verified to reproduce the chargino branching ratio results given in
Ref. \cite{branchingratios}.

Effects of initial state radiation (ISR) are not included in our
simulation. Hadronization and detector effects are crudely simulated by
smearing quark parton and lepton energies.  The detector resolutions
currently available at the ALEPH detector at LEP are \cite{ALEPH}

\be
\sigma^{\rm had}_E / E = 80 \%/\sqrt{E}$ \quad and \quad
$\sigma^{\rm e.m.}_E / E = 19 \%/\sqrt{E} \ ,
\ee
where $E$ is in GeV. In the Monte Carlo simulation, we smear the lepton
parton energies by the leptonic resolution. For the typical energies of
our simulation, the resulting leptonic resolutions are numerically a
good approximation to those that will be achieved for both muons and
electrons by current LEP detectors.  We also smear the hadronic parton
energies by the hadronic resolution.  By doing so, we implicitly assume
that the quark jet energy is measured by the hadronic calorimeter only.
In fact, however, quark jets are detected by a combination of the
tracking chamber, the electromagnetic calorimeter, and the hadronic
calorimeter, and, in particular, the addition of tracking chamber
measurements can improve the jet energy resolution substantially. This
improvement was studied in Ref.~\cite{Miyamoto} for $W$ mass resolution
at a $\sqrt{s} = 500 \gev$ $\epem$ collider.  For a detector with
energy resolutions $\sigma^{\rm had}_E / E = 50 \%/\sqrt{E} + 2\%$ and
$\sigma^{\rm e.m.}_E / E = 10 \%/\sqrt{E} + 2\%$, the resolution of
$M_W$, as measured by the dijet mass, was found to improve by 33\% from
4.1 GeV to 2.7 GeV when the tracking chamber measurements were
included. Similar studies for the L3 detector at LEP have shown that
when the momentum measurement from the central tracking detector is
included, the resolution for the total energy in hadronic events
improves by about 20\% from 9.2 GeV to 7.6 GeV \cite{L3}. To simply
account for the improvements from tracking chamber measurements, we
will reduce our resolutions for $m_{jj}$ and $E_{jj}$ endpoint
determinations by 25\% by hand in the case studies presented below.

To study the observables presented in Sec.~\ref{sec:observables}, we
must find the $m_{jj}$ and $E_{jj}$ distributions in $2j+l$ chargino
events and determine two of the partial cross sections, including that
of the mixed mode.  Three of these measurements can be made with $Y$
mode events, the subset of $2j+l$ events in which the hadrons do not
come from a tau lepton, since the $Y$ mode cross section can be
converted to a mixed mode cross section (under the assumption that
lepton universality holds).  It is therefore important to include
realistic cuts that isolate the $Y$ mode chargino events.  We will now
show that $2j+l$ events may be easily resolved.  In
Sec.~\ref{sec:studies} we will see that the $Y$ mode events in this
sample may be isolated by simply eliminating events with low invariant
hadronic mass.

The $2j+l$ mode is the most promising for chargino discovery and has
been carefully studied \cite{Aachen,Chen,Grvz2}. In Ref.~\cite{Grvz2}
cuts have been designed for the parameters $\sqrt{s} = 175\gev$,
$\mu=-100 \gev$, $\tanb=2$, $\mchargino=80\gev$, and $\mLSP=$ 20--60
GeV.  These cuts include the following requirements:

\noindent 1) The missing transverse momentum $\mpt$ is required to be
greater than 10 GeV.

\noindent 2) The event must contain an isolated electron or muon with a
momentum larger than 5 GeV and with no hadronic activity within
$30^{\circ}$.

\noindent 3) The squared missing mass $m_{\rm missing}^2$ must be
greater than $4000 \gev^2$.

\noindent 4) The hadronic system mass $m_{jj}$ must be less than 45 GeV.

\noindent 5) Under the assumption that the missing momentum in the
event is due to a ``neutrino'', the mass of the lepton-``neutrino''
system is required to be less than 70 GeV.  This removes most $WW$
events.  Actually, one can do even better than this, since it is
possible to allow for two unobserved massless particles, one an
undetected ISR photon along the beam axis and the other a missing
``neutrino'', and still determine all of the momenta in the event. In
this case, forcing the lepton-``neutrino'' invariant mass to be less
than 70 GeV removes many $WW\gamma_{ISR}$ events as well.

As shown in Ref.~\cite{Grvz2}, cuts 1 and 2 reduce the $WW$ background
to 2.8 pb.  Cuts 3--5 are specifically designed to reduce this
background further.  After additionally imposing cuts 3, 4, and 5, the
$WW$ background is reduced to 180 fb, 17 fb, and 7 fb, respectively.
After all cuts, the other standard model backgrounds contribute only 2
fb. Applying these same cuts to a chargino sample, we have found that
typically the mixed mode is reduced by about 25--75\% after cut 4, but
40\% of these are eliminated by cut 5.  Although the additional 40\%
loss in statistics is not extremely large, typically the signal to
background ratio is greater than or of order 50 after cut 4 and the
background is already negligible. Thus, of the five cuts listed, we
will use cuts 1--4 and ignore cut 5 in our analysis. This leaves the
standard model background at approximately 20 fb.

As we do not include the effects of hadronization, we will also require
that each final state quark parton have energy greater than 5 GeV so
that its hadronization products are detected.

In addition to standard model backgrounds, there may be supersymmetric
backgrounds that will need to be distinguished from chargino pairs.  In
particular, as noted in Sec.~\ref{sec:susyparam}, if charginos are
produced, neutralino production is almost certainly allowed. It is
first worth noting that the neutralino background is highly suppressed
in a significant portion of parameter space \cite{neutralino}.
Neutralinos are produced through a $Z$ annihilation diagram and
$t$-channel selectron exchange.  However, as the $Z$ couples only to
the Higgsino components of the neutralinos, neutralino production is
suppressed in the gaugino region unless the selectron mass is low.  It
may happen, then, that neutralino production, though kinematically
allowed, is a negligible background to chargino production.  For other
regions of parameter space, it should still be possible to separate the
neutralino and chargino events.  Production of $\chn_1\chn_2$ is not a
significant background to $Y$ mode chargino events, because these
neutralino events produce exactly one isolated charged lepton only when
$\chn_2\rightarrow \chn_1\tau\bar{\tau}$, and one $\tau$ decays
leptonically and the other hadronically.  These should be easily
distinguished from $Y$ mode chargino events, based on the small
invariant mass of the hadrons.   Production of $\chn_2\chn_2$ may be
problematic if one neutralino decays to $\chn_1\tau\bar{\tau}$ while
the other decays to $\chn_1 q\bar{q}$, which can lead to an isolated
lepton and an assortment of hadrons.  However, even if the hadrons
cannot be resolved into three isolated jets, kinematics often
distinguish these events from chargino events, since the total hadronic
energy or invariant mass may exceed what is allowed in
$\chc_1\rarr\chn_1 q\bar{q}$.  In any case, the number of these events
is usually very small.  Both types of neutralino events may be
backgrounds for the purely leptonic chargino events, and $\chn_2\chn_2$
may also be a background for the four-jet events. However, neutralino
events do not produce significant numbers of $e \mu$ events, nor do
they produce $\bar{l}\, l$ events with $m_{\bar{l}\, l} >
m_{\chn_2}-\mLSP$, and these may therefore be identified as chargino
events. Assuming lepton universality, one can then determine both the
total leptonic and total hadronic cross section from chargino pairs.

If cascade decays occur with a substantial branching fraction, they may
also make it difficult to isolate the chargino signal.  Such decays are
prominent, for example, in certain regions with $M_2 \sim |\mu| \sim
\mw$, where the mass separation between the $\chargino$ and the
$\chn_2$ is large, and where $\chn_1\chn_3$ production may be possible.
We assumed in Sec.~\ref{sec:susyparam} that cascade decays of the
chargino itself have a small branching fraction; for points where this
is not true our analysis must be modified.  For $m_{\chn_2} >
m_{\chargino}$ the decay of the $\chn_2$ through a chargino, which then
decays to an LSP, can be prominent, and can be a background to chargino
events.  The $\chn_3$ may also undergo cascade decays.  To distinguish
these cases, and to isolate the chargino signal, one should vary the
beam energy and make use of the fact that each signal has a unique
energy dependence.  Our Monte Carlo simulation does not include these
supersymmetric backgrounds, and we have not studied the effectiveness
of changing the beam energy, but we will assume that an energy scan
will allow the separation of the chargino signal from cascade decays.
In any event, as the SUSY parameters become roughly known, it is
possible that improved cuts could be devised to efficiently separate
the chargino and neutralino signals.

Approximate knowledge of the SUSY parameters may also be relevant for
the isolation of the chargino signal from standard model backgrounds.
The most obvious example is the possibility of a chargino with mass
less than $M_W$, in which case one could work below the $WW$ threshold,
dispense with cuts 3--5, and increase the number of signal events by
approximately a factor of 2. However, for simplicity in this
exploratory study, we will use cuts 1--4 and the requirement on quark
energies in all regions of parameter space.

\section{Case Studies}
\label{sec:studies}

In this section, we present a number of case studies at representative
points in SUSY parameter space. In the first subsection, we discuss the
general procedure that will be used to find the allowed regions of SUSY
parameter space.  We also describe the way in which our results will be
presented graphically. In the remaining subsections, we consider points
in the gaugino, Higgsino, and mixed regions, and determine for each
case how well the observables may be measured and what bounds on
underlying SUSY parameters may be obtained. Throughout this section, we
present results for $\sqrt{s}=190 \gev$ and an event sample of $1 \ifb$.

\subsection{Strategy for Finding Allowed Parameter Space}
\label{sec:strategy}

After the observables $\mchargino$, $\mLSP$, $\stot$, and $B_l$ are
measured, one must determine how the SUSY parameter space is
restricted.  As the parameter space is six-dimensional, it is important
to outline a method by which such restrictions are easily applied and
understood.

The dependence of $\mchargino$ on only three SUSY parameters allows us
a simple starting point. First, consider the three-dimensional space
$(\mu, M_2, \tanb)$.  A point $\cal P$ in this space survives the
chargino mass measurement if it predicts the chargino mass correctly,
within experimental uncertainties. The allowed region is then confined
to two thin sheets which we will label as $\cal S$, one with $\mu<0$
and another with $\mu>0$. This is shown schematically in
Fig.~\ref{fig:sheets}.

To display our results, it will be necessary to plot contours in the
allowed region, and it will therefore be convenient to flatten the two
sheets $\cal S$ into a plane $\cal T$ with the coordinate transformation

\be
(\mu, M_2, \tanb ) \in {\cal S} \rightarrow \left(\alpha\equiv \arctan
\frac{M_2}{\mu}, \tanb \right) \in {\cal T} \ ,
\ee
as shown in Fig.~\ref{fig:sheets}. Since the sheets are not infinitely
thin, a short segment of points in $\cal S$ is projected into every
point in $\cal T$.  The far gaugino regions are then transformed to the
areas with $\alpha \approx 0^\circ , 180^\circ$, and the far Higgsino
regions now correspond to the region with $\alpha \approx 90^\circ$.
The symmetry $\mu\leftrightarrow -\mu$ for $\tanb\rarr\infty$ implies
that, at large $\tanb$, observables at $\alpha$ are nearly equal to
those at $180^\circ-\alpha$.

The allowed region is restricted further by the other measurements. The
LSP mass $\mLSP$ is a function of $\mu$, $M_2$, $\tanb$, and $M_1$, and
so the $\mLSP$ measurement limits $M_1$ to a certain range.  To
graphically represent this restriction of $M_1$, or equivalently,
$M_1/M_2$, we will do the following.  For a point ${\cal P} =
(\alpha,\tanb) \in {\cal T}$, we find all parameters $(\mu,
M_2=\mu\tan\alpha, \tanb, M_1/M_2)$ such that the corresponding values
of $\mchargino$ and $\mLSP$ are within the experimental limits. The
allowed values of $M_1/M_2$ will lie in some range $(M_1/M_2)_{\rm min}
< M_1/M_2 < (M_1/M_2)_{\rm max}$.  To display this range, we will plot
contours in $\cal T$ of $(M_1/M_2)_{\rm min}$ and $(M_1/M_2)_{\rm
max}$. If no value for $M_1$ in the range given in Eq.~(\ref{bounds})
leads to the correct $\mLSP$, the point ${\cal P}$ is excluded.

In a similar manner, the measurement of $\stot ( \mu, M_2, \tanb,
\mslep)$ will limit the allowed range of $\mslep$, and this range can
be represented in contour plots of $(\mslep)_{\rm min}$ and
$(\mslep)_{\rm max}$. If no value of $\mslep$ gives the correct
$\stot$, ${\cal P}$ is not allowed. Finally, given $\cal P$ and the
ranges of $M_1$ and $\mslep$ as determined above, the measured value of
$B_l$ restricts the range of $\msq$ to a certain range, and we will
also present contours of $(\msq)_{\rm min}$ and $(\msq)_{\rm max}$.  If
no value of $\msq$ yields the appropriate $B_l$, then the point $\cal
P$ is excluded.  In principle, measurement of $\afbjj$ may also be used
to limit the allowed region, but we defer discussion of this observable
to the individual case studies below.

The remaining points ${\cal P} = (\alpha,\tanb)\in {\cal T}$, for which
there exist parameters $(\mu, \ M_2=\mu\tan\alpha,\
\tanb,M_1,\mslep,\msq)$ that are consistent with all measurements, form
the allowed region.  We will plot this region in the $\cal T$ plane.
{}From this plot, the allowed ranges of $\rhoplus$ and $\tanb$ may be
quickly determined. The ranges of and correlations between the other
variables may be found from the contour plots of their minimum and
maximum allowed values.

\subsection{Gaugino Region}
\label{subsec:gaugino}

The first case we consider is a set of SUSY parameters in the gaugino
region with gaugino mass unification.  We choose

\be
(\mu, M_2, \tanb, M_1/M_2, \mslep, \msq) = (-400,75,4,0.5,200,300).
\ee
With these parameters, $\alpha = 169^\circ$, and the underlying values
and ranges of the most important quantities are

\be\label{gvalues}
\begin{array}{rcl}
\mchargino &=& 79.6 \gev \\
\mLSP      &=& 39.1 \gev \\
\stot      &=& 1.16 {\rm R} = 3200 \fb \\
\rhoplus   &=& 1.00 \\
\rhozero   &=& 0.99 \\
B_l        &=& 0.42 \\
m_{jj}     &<& 40.5 \gev \\
16.4 \gev  &<& E_{jj} < 55.8 \gev \ .
\end{array}
\ee
Since $\mchargino \approx M_W$, it is not possible to work below the
$WW$ threshold, and the cuts for $2j+l$ events described in
Sec.~\ref{sec:simulation} are likely to be nearly ideal.  Note that the
sneutrino mass has been taken near the low end of the range.  This
value leads to substantial destructive interference in the production
amplitude; higher values for $\msnu$ would give considerably larger
cross sections.

With an integrated luminosity of $1 \ifb$, there are 3203 chargino
events, and the Monte Carlo simulation yields 1493 mixed mode events.
Some of these events include hadronically-decaying $\tau$ leptons, but
the rest of them are $Y$ mode events, as defined in
Sec.~\ref{sec:susyparam}.  In the Monte Carlo simulation we are left
with 1184 $Y$ events, of which 889 (75\%) survive the cuts described in
Sec.~\ref{sec:simulation}.  In addition to these $Y$ mode events, some
leptonic mode chargino events with hadronically-decaying $\tau$ leptons
will also pass the cuts.  However, hadrons resulting from $\tau$ decays
are highly collimated with invariant mass less than $m_{\tau}$. The
$m_{jj}$ spectrum for the 889 $Y$ events is shown in
Fig.~\ref{fig:gmjjEjj}a. Clearly, very few $Y$ mode events have dijet
masses consistent with $\tau$ decays, and $Y$ events should be easily
separated on this basis. We will therefore assume that we have an event
sample of $Y$ events that is virtually free of background and may be
used to determine the values of chargino event observables.

We may now determine the masses $\mchargino$ and $\mLSP$ from the
endpoints of dijet mass and energy distributions, as discussed in
Sec.~\ref{sec:observables}. These distributions are given in
Fig.~\ref{fig:gmjjEjj}.  We see that finite detector resolution effects
cause the spectra to have tails that exceed the theoretical limits, but
despite this, the endpoints are fairly sharp. The $\mjjmax$ endpoint
almost certainly lies within a 8 GeV range, and we therefore estimate
its 1$\sigma$ error to be 2 GeV. Similarly, we estimate that the
1$\sigma$ error for the maximum endpoint of $E_{jj}$ is 3 GeV.  As
noted in Sec.~\ref{sec:simulation}, these resolutions are expected to
improve with the addition of tracking chamber momentum measurements
\cite{Miyamoto,L3}, and we therefore take the actual resolutions to be
reduced by 25\% to $\Delta\mjjmax = 1.5 \gev$ and $\Delta\Ejjmax = 2.3
\gev$.  (In the next subsection we will examine the effect on our
results of increasing these uncertainties.)  We must now propagate
these uncertainties into the determinations of the underlying masses.
The relevant formulae for the uncertainty calculations are collected in
the appendix. From Eqs.~(\ref{massesfrommE}) and (\ref{masserror1}) we
find

\be
\Delta\mchargino= 2.5 \gev  \quad {\rm and } \quad
\Delta\mLSP= 2.2 \gev \ .
\ee
For simplicity, we have assumed that the central values for endpoint
measurements are their underlying physical values. We note also that,
although $\Ejjmin$ provides a useful cross check and may also improve
the mass determinations, we will not use it here.

Next we must determine $\stot$ and $B_l$ and the uncertainties in their
measurements. To do this, as discussed in Sec.~\ref{sec:observables},
we must measure at least two of the three partial cross sections
$\sigm$, $\sigh$, and $\sigl$. The partial cross section $\sigm$ is
always one of the two largest, and we will consider this mode in
detail.  As noted above, we do not measure $\sigm$ directly, but rather
$\sigy$.  Under the assumptions of Sec.~\ref{sec:susyparam}, which
imply lepton universality, $\sigy = \frac{2.0+0.4}{3.0} \sigm =
\frac{4}{5}\sigm$ and $\frac{\Delta \sigy}{\sigy} = \frac{\Delta
\sigm}{\sigm}$.  We will then assume that similar errors, in a sense to
be defined precisely below, may be obtained for the hadronic mode. With
these assumptions, we then find the uncertainties of $\stot$ and $B_l$.
(The consistency of our assumption of lepton universality may be
checked by verifying that, for example, $B(\chcp_1\chcm_1 \rarr 2j+e) =
B(\chcp_1\chcm_1 \rarr 2j+\mu)$.  Furthermore, if additional branching
ratios can be measured, one may determine directly whether lepton
universality indeed holds.)

We must now determine the uncertainty for measurements of $\sigy$.  For
any mode $i$, the cross section and fractional uncertainty are

\be\label{sigmai}
\sigi = N_i \eta_i^{-1} {\cal L}^{-1}
\ee
and

\be\label{deltasigmai}
\frac{\Delta\sigi}{\sigi}
= \left[ \left( \frac{\Delta N_i}{N_i} \right)^2
+ \left( \frac{\Delta\eta_i}{\eta_i} \right)^2
+ \left( \frac{\Delta{\cal L}}{\cal L} \right)^2 \right]^{\half} \ ,
\ee
where $N_i$ is the number of $i$ mode events passing the cuts, $\eta_i$
is the efficiency of the cuts for $i$ mode events, and $\cal L$ is the
collider luminosity. The number of $Y$ mode events passing the cuts is
889, so $\Delta N_Y / N_Y = 1/\sqrt{N_Y} = 3.4\%$. The efficiency
$\eta_Y$ is not known and, in principle, depends on all the SUSY
parameters that we are trying to determine.  However, by running Monte
Carlo simulations for many points in SUSY parameter space with the
measured $\mchargino$ and $\mLSP$, we can determine how much the
efficiency varies throughout the allowed parameter space.  We have done
this for points in the gaugino, Higgsino, and mixed regions, for both
positive and negative $\mu$, various $\tanb$, and $(\mslep, \msq)$ =
(100, 150), (200, 300), and (500, 700), with the sole restriction being
that these points give $\mchargino\approx 80 \gev$ and $\mLSP\approx 39
\gev$. For all of these cases, the efficiency of the cuts varies only
between 70\% and 77\%.  Thus, the cut efficiency is determined
primarily by kinematics and varies only slightly for fixed $\mchargino$
and $\mLSP$. We take the efficiency to be $\eta_Y = 73.5\%$ and its
variation to be $\Delta\eta_Y = 3.5\% $. The uncertainty in the
luminosity, which at LEP I is $\Delta{\cal L}/{\cal L} = 0.3\%$, and
which is not expected to increase substantially for LEP II
\cite{Marta}, is much smaller than the other errors.  Substituting
these values into Eq.~(\ref{deltasigmai}), we find that $\Delta\sigy
/\sigy = 5.8\%$.

Although we do not have specific cuts for the hadronic and leptonic
modes, we will assume that cuts with similar $\eta$ and $\Delta\eta$
may be devised for at least one of the other modes.  Such an assumption
is certainly not to be taken for granted as the other two modes have
large backgrounds.  In the four jet mode, it may be possible to reduce
backgrounds substantially by demanding that no pairing of jets yields
two dijet masses consistent with $M_W$ or $M_Z$.  The leptonic mode is
plagued by an irreducible background from $W$ pair production
\cite{Chen}.  As can be seen from Eq.~(\ref{deltasigmai}), the best
cuts for the purposes of this study are those that balance uniformity
(low $\frac{\Delta\eta_i}{\eta_i}$) with background suppression and
efficiency (low $\frac{\Delta N_i}{N_i}$).  It is clear that detailed
studies of cuts for the $4j$ and $2l$ events are necessary for future
work in this area.  For this study, however, we will calculate the
fractional uncertainty in $\sigh$ assuming that cuts may be devised
with values of $\eta$ and $\Delta\eta$ similar to those obtained in the
mixed mode. It is important to note that significantly worse values of
$\eta_{\rm hadronic}$ need not change our main results dramatically. We
will demonstrate this explicitly in the following subsection, where
results are presented for lower values of $\eta_{\rm hadronic}$. We
will assume in our case studies that the errors for the leptonic mode
are larger than those for the hadronic mode, and so we will use only
the mixed and hadronic modes for our determinations of $\stot$ and
$B_l$.  In a complete analysis the leptonic mode should be combined
with the others to further constrain the determinations of $\stot$ and
$B_l$.

The Monte Carlo simulation yields 1095 hadronic mode events. The
assumption $\eta_{\rm hadronic} = \eta_Y = 73.5\%$ implies $N_{\rm
hadronic} = 804$, and assuming also that $\Delta\eta_{\rm hadronic} =
3.5\%$, we find that $\Delta\sigh /\sigh = 5.9\%$. We may now proceed
to determine the uncertainties in $\stot$ and $B_l$.  Using the
formulae in the appendix, we find

\be
\Delta\stot /\stot = 5.0\% \quad {\rm and} \quad
\Delta B_l /B_l = 4.8\% \ .
\ee

We have now determined the uncertainties with which the four
observables may be measured and can apply the strategy outlined in
Sec.~\ref{sec:strategy} to determine the allowed region in SUSY
parameter space.  To recapitulate, the $\mchargino$ measurement
restricts the $(\mu, M_2, \tanb)$ space to two thin sheets, one with
$\mu<0$ and another with $\mu>0$. These sheets are then flattened into
the $(\alpha, \tanb )$ plane, where $\alpha$ is the angle given by
$\tan\alpha = M_2/\mu $.  The far gaugino regions are at $\alpha
\approx 0^{\circ}, 180^{\circ}$, while the far Higgsino region lies
near $\alpha \approx 90^{\circ}$.

For any given point $\cal P$ in the $(\alpha, \tanb )$ plane, we
determine values of $M_1/M_2$ that give the measured value of $\mLSP$
within 1$\sigma$ bounds. In general, there will be an allowed range for
$M_1>0$ and another for $M_1<0$. These are distinct branches, as there
are no symmetries connecting positive and negative $M_1$. For now, let
us investigate the $M_1>0$ possibility only.

In Fig.~\ref{fig:gM1range} we plot constant contours of the minimum and
maximum allowed values of $M_1/M_2$ in the $(\alpha, \tanb )$ plane.
The approximate symmetry $\mu \leftrightarrow -\mu$ for large $\tanb$
is already in evidence at $\tanb = 10$, and the contour lines are
approximately independent of $\tanb$ above this value. If no value of
$M_1/M_2$ gives the correct $\mLSP$, the point $(\alpha,\tanb)$ is
excluded, as happens in the small cross-hatched area in the
$\alpha>90^{\circ}$ ($\mu<0$) mixed region with $\tanb\approx 1$, where
it is not possible to raise $M_1/M_2$ sufficiently to produce the
required $\mLSP$.  For the areas of the plane that are not excluded,
the range of allowed $M_1/M_2$ values is quite restricted. In the far
gaugino region, the allowed range of $M_1/M_2$ is roughly $0.45 \alt
M_1/M_2 \alt 0.55$ and is centered around 0.5, as expected. In the
Higgsino region, for decreasing $\rhoplus$, $M_1/M_2$ drops to zero.
This is easily understood, since, as we approach the far Higgsino
region, for which $M_2$ is large, $M_1$ must remain roughly constant at
$M_1 \sim 40 \gev$ to accommodate the neutralino mass $\mLSP \approx 40
\gev$. We also see that, for $\tanb \agt 2.5$, only values of $M_1/M_2$
less than 0.6 are allowed.

We now determine the values of $\mslep$ that give the observed $\stot$
within 1$\sigma$ bounds. In Fig.~\ref{fig:gmsnurange} we plot the
minimum and maximum allowed values of $\mslep$ for points in the
$(\alpha, \tanb )$ plane. In the far gaugino region, the slepton mass
range is $180 \gev \alt \mslep \alt 220 \gev$. The bounds are quite
stringent because the cross section is very sensitive to $\msnu$ in the
gaugino region, as was shown in Fig.~\ref{fig:sigmasnu}. As one moves
from the gaugino region to the Higgsino region, the cross section for a
fixed $\mslep$ decreases, and therefore $\mslep$ increases to
compensate.  In addition, the cross sections become less sensitive to
$\mslep$, and the uncertainty for the $\mslep$ determination grows.
Finally, at a certain point, $\mslep$ cannot be large enough to prevent
the cross section from dipping below the measured value, and thus the
far Higgsino region is excluded.  The $\stot$ measurement alone is
therefore enough to exclude the cross-hatched region of the $(\alpha,
\tanb)$ plane in Fig.~\ref{fig:gmsnurange}.

For the remaining allowed regions of the $(\alpha, \tanb)$ plane, we
use the determined ranges of $M_1$ and $\mslep$ to find the values of
$\msq$ that give the correct $B_l$ within 1$\sigma$. Contours of
constant $(\msq)_{\rm min}$ and $(\msq)_{\rm max}$ are plotted in
Fig.~\ref{fig:gmsqrange}, where the $\alpha > 90^{\circ}$ ($\mu < 0$)
gaugino region has been enlarged. (Similar features are seen in the
$\alpha < 90^{\circ}$ ($\mu > 0$) gaugino region.)  As one approaches
the far gaugino region, the leptonic branching fraction grows for fixed
$\msq$, and the maximum allowed squark mass $(\msq)_{\rm max}$
decreases. At some point, no squark mass greater than 150 GeV is
allowed, and the hatched region is therefore excluded.  Since $B_l$
grows more quickly for large $\tanb$, as predicted by
Eqs.~(\ref{bllowtanb}) and (\ref{blhightanb}), the excluded region is
larger for high $\tanb$. Conversely, if one moves from the far gaugino
region to the gaugino region, $B_l$ drops, and to stay within the
1$\sigma$ bounds on $B_l$, $(\msq)_{\rm min}$ grows. At a certain
point, no $\msq$ is large enough to accommodate the measured $B_l$, and
so the cross-hatched region is also excluded. The resulting allowed
regions are shown in Fig.~\ref{fig:gallowed}, where the cross-hatched
regions are excluded. We see immediately that the allowed regions lie
completely in the gaugino region.  (A subtle point should be mentioned
here.  The $B_l$ measurement not only constrains the allowed region in
the $(\alpha, \tanb)$ plane, but also, for a fixed point in the allowed
region, further limits the acceptable values of $M_1$ and $\mslep$.
Therefore the allowed ranges may be somewhat reduced from those in
Figs.~\ref{fig:gM1range} and \ref{fig:gmsnurange}. The effect is
typically small, however, and so we do not present updated figures for
$M_1$ and $\mslep$ with the $B_l$ constraint imposed.  Nonetheless, the
full $B_l$ constraint is included in the results presented below.)

It is evident from the figures that the four measurements constrain the
parameter space significantly, and the allowed ranges of the SUSY
parameters are highly correlated. It is also useful to determine the
global bounds that may be placed on the various quantities of interest,
independent of their correlations. To determine the allowed ranges of
these quantities, we pick points randomly in the allowed volume, and
verify graphically that enough points have been picked to adequately
sample the region.  We find the following global bounds:

\be\label{grangespos}
\begin{array}{rcccl}
0.97    &<& \rhoplus        &<& 1.00 \\
0.97    &<& \rhozero        &<& 1.00 \\
180\gev &<& \mslep          &<& 225 \gev \\
0.43    &<& \frac{M_1}{M_2} &<& 0.58 \\
-1\tev   <  \mu              < -290 \gev
&\quad& {\rm or} &\quad&
300 \gev <  \mu              <  1 \tev \\
63\gev  &<& M_2             &<& 93\gev \\
1       &<& \tanb           &<& 50 \\
150\gev &<& \msq            &<& 1\tev \ .
\end{array}
\ee
To understand the confidence level of these bounds, recall that the
uncertainties in observables we have used are one standard deviation.
The allowed region consists of all points in parameter space for which
the central values of all observables are within 1$\sigma$ of the
underlying physical values. Typically, the extremes of the quantities
given in Eq.~(\ref{grangespos}) are reached in corners of the allowed
region for which more than one of the observables deviates by 1$\sigma$.

These bounds follow from the assumption that $M_1>0$.  For the case of
$M_1<0$, the resulting bounds are only very slightly weaker.  For
general $M_1$, we find that for the gaugino region point we have
chosen, the global bounds are

\be\label{grangesall}
\begin{array}{rcccl}
0.97    &<& \rhoplus        &<& 1.00 \\
0.97    &<& \rhozero        &<& 1.00 \\
179\gev &<& \mslep          &<& 227 \gev \\
-0.61    <  \frac{M_1}{M_2}  <  -0.45
&\quad& {\rm or} &\quad&
0.43     <  \frac{M_1}{M_2}  <  .58 \\
-1\tev   <  \mu              < -290 \gev
&\quad& {\rm or} &\quad&
300 \gev <  \mu              < 1 \tev \\
63\gev  &<& M_2             &<& 93\gev \\
1       &<& \tanb           &<& 50, \\
150\gev &<& \msq            &<& 1\tev \ .
\end{array}
\ee
The allowed regions in Fig.~\ref{fig:gallowed} are virtually unchanged
when negative $M_1$ is included.

Even though correlations between variables are ignored, the global
bounds of Eq.~(\ref{grangesall}) have interesting implications. The
gaugino content has been tightly constrained to be greater than 0.9,
which supports the LSP as a dark matter candidate. For the case of
$M_1>0$, the ratio $M_1/M_2$ has been determined to be compatible with
grand unification to within approximately 15\%. There is, however, an
allowed range of negative $M_1$. (In general, it is very difficult to
exclude negative $M_1$ with the four observables we have explored.) The
bound on $\mslep$ is strong, as a result of the large destructive
effect of the electron sneutrino on the total cross section. In many
models the sneutrino is the next lightest observable SUSY particle, and
this bound provides an important guide for future sparticle searches.
Finally, $\tanb$ is unrestricted, and there is no global bound on
$\msq$ --- the squark mass may lie anywhere in the range we have
considered.  However, as seen in Fig.~\ref{fig:gmsqrange}, at a given
point in the $(\alpha, \tanb)$ plane, the bounds on $\msq$ may be quite
strong.

We now turn to $\afbchi$ and $\afbjj$, the forward-backward asymmetries
discussed in Sec.~\ref{sec:observables}.  We remind the reader that the
former, while unobservable, depends only on the production amplitude,
while the latter is observable but depends on the decay vertices as
well.  In Fig.~\ref{fig:gafb}, we plot $\afbjj$ vs. $\afbchi$ for a
number of points in the allowed region.  The value of $\afbjj$ for our
case study, measured from the $Y$ mode events that pass the cuts, is
given by the solid line.  Its 1$\sigma$ deviation, as determined from
Eq.~(\ref{afberror}), is given by the dashed lines.

Because the previous four observables have already restricted the
allowed region to a small volume in the gaugino region with a light
sneutrino, $\afbchi$ is limited to the fairly narrow range $0.12 \alt
\afbchi \alt 0.21$.  As discussed in Sec.~\ref{sec:observables},
$\afbchi$ and $\stot$ are the only production quantities with much
resolving power, so the fact that $\afbchi$ is limited to a small range
is evidence that we have already obtained nearly all of the information
contained in the production amplitude.

We also see that the correlation between $\afbjj$ and $\afbchi$ is
weak, and that $\afbjj$ lies in a much broader range,
$-0.06\alt\afbjj\alt 0.3$. Clearly the decay amplitude plays a crucial
role in determining the value of $\afbjj$, and the large variation in
$\afbjj$ is an indication that detailed studies of chargino decays may
improve the bounds on parameter space and tighten the correlations
between variables. By running Monte Carlo simulations for a large
number of points that densely populate the allowed region, one could
presumably form a detailed picture of the regions that may be excluded
on the basis of $\afbjj$.  However, as our goal in this study is to
study chargino events analytically, we will not discuss this
possibility further. Still, even from our sparse sampling of the
allowed region it is possible to draw some tentative conclusions. For
example, in Fig.~\ref{fig:gafb}, points with positive $\mu$ have been
marked with open circles, and those with negative $\mu$ have been
marked with filled circles. We see that all of the points with $\mu >
0$ may be excluded based on the $\afbjj$ measurement, which suggests
that the $\mu>0$ portion of the allowed region may be eliminated by the
$\afbjj$ measurement.

\subsection{Variations in $\frac{M_1}{M_2}$ and Experimental
Assumptions}

We will now briefly explore two simple variations on the previous case
study.  In that example, we assumed gaugino mass unification and saw
that it could be verified to 15\% (assuming $M_1>0$).  In our first
variation below, we take $M_1/M_2= 0.7$ and find how strongly the
gaugino mass unification condition may be disfavored.  In the second
variation, we determine the impact of more pessimistic assumptions
about detector resolutions and backgrounds.  As these are only slight
variations on the previous example, few new features appear in the
analysis, and we will only present a few intermediate results and the
final conclusions.

We begin by taking the parameters

\be
(\mu, M_2, \tanb, M_1/M_2, \mslep, \msq) = (-400,75,4,0.7,200,300) \ ,
\ee
where the only change from the previous example is that we choose
$M_1/M_2= 0.7$.  Many of the basic quantities remain the same, but we
list them all for convenience:

\be\label{gvalues2}
\begin{array}{rcl}
\mchargino &=& 79.6 \gev \\
\mLSP      &=& 53.8 \gev \\
\stot      &=& 1.16 {\rm R} = 3202 \fb \\
\rhoplus   &=& 1.00 \\
\rhozero   &=& 0.99 \\
B_l        &=& 0.40 \\
m_{jj}     &<& 25.7 \gev \\
11.7 \gev  &<& E_{jj} < 39.9 \gev \ .
\end{array}
\ee

Of the 3203 chargino events, the Monte Carlo simulation yields 1528
mixed mode events.  1210 of these are $Y$ mode events, and 715 (59\%)
of these survive the cuts.  Because the LSP is heavier, the ranges of
the dijet mass and energy are smaller than in the original example.
{}From plots of the dijet mass and energy distributions similar to
those previously presented, we estimate that both $\mjjmax$ and
$\Ejjmax$ may be determined to 2 GeV.  Including the 25\% reduction in
these uncertainties from the tracking chamber measurements, we find
that the chargino and LSP masses are determined to

\be
\Delta\mchargino= 2.9 \gev  \quad {\rm and } \quad
\Delta\mLSP= 2.0 \gev \ .
\ee

The efficiency $\eta_Y$ is found, as in the previous example, by
running Monte Carlo simulations for a large number of points in SUSY
parameter space with the correct $\mchargino$ and $\mLSP$.  We find
once again that the efficiency is principally determined by kinematics
and lies in the range 54--62\%.  We take $\eta_Y = 58\%$ and
$\Delta\eta_Y = 4 \%$, and assume similar values for the hadronic mode.
Applying Eqs.~(\ref{deltasigmai}) and (\ref{stotblerr}), we find the
uncertainties

\be
\Delta\stot /\stot = 6.6\% \quad {\rm and} \quad
\Delta B_l /B_l = 6.4\% \ .
\ee

We now apply these measurements to the SUSY parameter space.  By
randomly sampling SUSY parameter space with both signs of $M_1$, we
have found the following bounds in the allowed region:

\be\label{grangesall2}
\begin{array}{rcccl}
0.78    &<& \rhoplus        &<& 1.00 \\
0.84    &<& \rhozero        &<& 1.00 \\
175\gev &<& \mslep          &<& 233 \gev \\
-1.00    <  \frac{M_1}{M_2}  <  -0.38
&\quad& {\rm or} &\quad&
0.60     <  \frac{M_1}{M_2}  <  1.00 \\
-1\tev   <  \mu              < -188 \gev
&\quad& {\rm or} &\quad&
230 \gev <  \mu              < 1 \tev \\
54\gev  &<& M_2             &<& 119\gev \\
1       &<& \tanb           &<& 50, \\
150\gev &<& \msq            &<& 1\tev \ .
\end{array}
\ee
The bounds on $M_1/M_2$ disfavor the gaugino mass unification
hypothesis.  The other conclusions and bounds are slightly weakened
relative to the original gaugino case study, but we still obtain strong
bounds on $\mslep$ and find that the allowed region lies primarily in
the gaugino region.

We now study the effects of varying our experimental assumptions. In
the analysis above we have attempted to estimate the effects of finite
detector resolutions and of backgrounds in the hadronic mode. Detailed
studies and simulations are needed to significantly improve the
accuracy of these estimates.  We will show here, however, that most of
the global bounds presented in the previous section are robust and are
not altered greatly by assuming poorer experimental conditions.  First,
we modify the case study of Sec.~\ref{subsec:gaugino} by assuming that
the backgrounds to the hadronic mode are very large, and that the
optimal cuts have an efficiency $\eta_{\rm hadronic} = 18\%$, which is
1/4 of the value we took previously. We retain the estimate $\Delta\eta
= 3.5\%$.  The uncertainties in $\mchargino$ and $\mLSP$ remain the
same, but we now find

\be\label{poorexperrors}
\begin{array}{rcl}
\Delta\sigh /\sigh &=& 20.3\%  \\
\Delta\stot /\stot &=& 5.9\% \\
\Delta B_l /B_l &=& 12.2\% \ .
\end{array}
\ee
Because we have assumed a low $\eta_{\rm hadronic}$, the uncertainties
in $\sigh$ and $B_l$ are large.  However, the uncertainty in $\stot$ is
not strongly affected because $1-2 B_l$ is small for this case study
(see Eq.~(\ref{stotblerr})).

The analysis is identical to the previous example, so we skip the
intermediate steps and present the end result.  In the allowed region
with both positive and negative $M_1$ values, the bounds on selected
quantities are:

\be\label{grangesall3}
\begin{array}{rcccl}
0.88    &<& \rhoplus        &<& 1.00 \\
0.94    &<& \rhozero        &<& 1.00 \\
176\gev &<& \mslep          &<& 231 \gev \\
-1.00    <  \frac{M_1}{M_2}  <  -0.30
&\quad& {\rm or} &\quad&
0.42     <  \frac{M_1}{M_2}  <  0.80 \ .
\end{array}
\ee
We see that the bounds are for the most part only slightly weakened,
with the exception that the upper bound on positive $M_1/M_2$ is now
$0.8$, and the lower bound on negative $M_1/M_2$ is $-1.0$.

We have also investigated the implications of doubling the estimated
uncertainties on the dijet mass and energy endpoint determinations.  In
this variation, we take $\Delta\mjjmax = 4.0 \gev$ and $\Delta\Ejjmax =
6.0 \gev$.  In addition, we assume no improvement from tracking chamber
measurements, and retain the ``poor'' hadronic mode efficiency of
$\eta_{\rm hadronic} = 18 \%$.  The uncertainties in $\stot$ and $B_l$
are as in Eq.~(\ref{poorexperrors}), and the new uncertainties in
$\mchargino$ and $\mLSP$ are

\be
\Delta\mchargino= 6.7 \gev  \quad {\rm and } \quad
\Delta\mLSP= 5.8 \gev \ .
\ee

The resulting bounds are

\be\label{grangesall4}
\begin{array}{rcccl}
0.83    &<& \rhoplus        &<& 1.00 \\
0.91    &<& \rhozero        &<& 1.00 \\
162\gev &<& \mslep          &<& 274 \gev \\
-1.00    <  \frac{M_1}{M_2}  <  -0.26
&\quad& {\rm or} &\quad&
0.37     <  \frac{M_1}{M_2}  <  1.00 \ .
\end{array}
\ee
In this case, we find that the bounds $M_1/M_2$ are weak. However, we
find that we are still able to constrain the parameter space to the
gaugino region and can place an upper bound on the sneutrino mass of
274 GeV. Thus, at least in the gaugino region, where our global bounds
are expected to be the strongest, the limits on $\rho$ and $\mslep$ are
robust under variations in efficiency for the hadronic mode and in
detector resolution.

\subsection{Higgsino Region}

For the Higgsino region case study, we choose the parameters

\be\label{hparam}
(\mu, M_2, \tanb, M_1/M_2, \mslep, \msq) = (-75,250,4,0.5,200,300) \ .
\ee
For this point, the angle $\alpha = 107^\circ$, and

\be\label{hvalues}
\begin{array}{rcl}
\mchargino &=& 79.7 \gev \\
\mLSP      &=& 62.3 \gev \\
\stot      &=& 0.89 {\rm R} = 2450 \fb \\
\rhoplus   &=& 0.17 \\
\rhozero   &=& 0.087 \\
B_l        &=& 0.34 \\
m_{jj}     &<& 17.4 \gev \\
8.4  \gev  &<& E_{jj} < 28.6 \gev \ .
\end{array}
\ee

This point in parameter space is again fairly central in the accessible
band, with $\mchargino \approx \mw$.  It is possible to study points in
parameter space that are closer to the pure Higgsino limit
$\rhoplus=0$. However, as discussed in Sec.~\ref{sec:observables}, as
one increases $\rhoplus$ and $M_2$, $\chc_1$ and $\LSP$ become more
nearly degenerate, and the number of events with soft jets increases.
These events are eliminated by our requirement that jet energies be
greater than 5 GeV, and the resulting event sample is small.  The point
in parameter space given in Eq.~(\ref{hparam}) has been chosen to have
properties characteristic of the Higgsino region, without being so far
in the Higgsino region that low statistics become the primary concern.

Given a sample of $1 \ifb$, there are 2454 events of which 1119 are
mixed mode events.  Considering only $Y$ mode events, that is,
excluding the mixed events in which a $\tau$ decays hadronically, we
are left with 893 events, of which 287 (32\%) survive the cuts
described in Sec.~\ref{sec:simulation}.  The efficiency of the cuts is
lower than in the gaugino example because the smaller chargino--LSP
mass splitting leads to more events with soft jets. In addition, since
the LSPs are produced with lower velocities in the chargino rest frames
and are more back-to-back in the lab frame, more events are eliminated
by the $\mpt$ cut.

We must now determine the masses from the endpoints of the $m_{jj}$ and
$E_{jj}$ spectra, which are shown in Fig.~\ref{fig:hmjjEjj}. As in the
gaugino case, finite detector resolution effects smear the endpoints,
but we estimate that the maximum $m_{jj}$ endpoint almost certainly
lies within a 4 GeV range, and therefore we takes its 1$\sigma$ error
to be 1 GeV.  Similarly, we estimate that the 1$\sigma$ error for the
maximum endpoint of $E_{jj}$ is 2 GeV.  As noted in
Sec.~\ref{sec:simulation}, these resolutions are expected to improve
with the addition of tracking chamber momentum measurements, and we
therefore take the actual resolutions to be reduced by 25\% to
$\Delta{\mjjmax} = 0.75 \gev$ and $\Delta{\Ejjmax} = 1.5 \gev$. (Again,
we will not consider the lower $E_{jj}$ endpoint in the analysis here,
although we expect it to be useful at least as a cross check.) The
resulting uncertainties from Eqs.~(\ref{massesfrommE}) and
(\ref{masserror1}) for $\mchargino$ and $\mLSP$ are

\be
\Delta\mchargino = 3.0 \gev \quad {\rm and} \quad
\Delta\mLSP  = 2.6 \gev \ .
\ee

We now turn to the determination of $\stot$ and $B_l$. As in the
gaugino region, the accuracy of these determinations depends on the
statistical uncertainties and the variation of the cut efficiencies in
the subvolume of SUSY parameter space with the given $\mchargino$ and
$\mLSP$.  With 287 events, the statistical uncertainty is $\Delta N_Y/
N_Y = 5.9\%$.  To determine the efficiency of the cuts, we have run
Monte Carlo simulations for a wide range of representative points with
$\mchargino \approx 80\gev$ and $\mLSP \approx 62\gev$, and find that
the cut efficiencies are fairly uniform and in the range of 26--34\%.
We take the efficiency to be $\eta_Y = 30\%$ and its variation to be
$\Delta\eta_Y = 4\%$.  Combining these uncertainties as in the gaugino
case, we find that $\Delta\sigy /\sigy = 15\%$. Finally, we will assume
that the formulae in Eq.~(\ref{stotblerr}) using the hadronic cross
section are the ones with the smallest uncertainties. There are 1057
hadronic mode events in the Monte Carlo simulation. Again taking the
assumption that $\eta_{\rm hadronic} \approx \eta_Y$ and $\Delta
\eta_{\rm hadronic} \approx \Delta \eta_Y$, we find $\Delta\sigh /\sigh
= 14\%$, and, from Eq.~(\ref{stotblerr}),

\be
\Delta\stot /\stot = 11\% \quad {\rm and} \quad \Delta B_l /B_l = 14\%
\ .
\ee

We may now use these four measurements to find the allowed parameter
space.  As in the gaugino example, $\mchargino$ limits us to two thin
sheets in $(\mu, M_2, \tanb )$ space, and these are flattened into the
plane $(\alpha, \tanb)$. We then determine the allowed region as in the
gaugino case by applying the bounds on $\mLSP$, $\stot$, and $B_l$ to
determine ranges of $M_1$, $\mslep$ and $\msq$ for every point in the
plane.  We will proceed as in the gaugino region case, first
considering only $M_1>0$, and then including the possibility $M_1<0$ in
the final determination of the allowed region.

The allowed range of $M_1/M_2$ from the $\mLSP$ measurement is shown in
Fig.~\ref{fig:hM1range}.  As in Fig.~\ref{fig:gM1range} of the gaugino
example, a portion of the low $\tanb$, $\mu<0$, mixed region is
excluded, and the allowed ratio drops to zero in the far Higgsino
region.  Relative to the gaugino example, however, the central value of
the allowed range of $M_1/M_2$ is increased in the gaugino region
because now $\mLSP / \mchargino = 0.78$.  The gaugino regions, in which
$\rhoplus, \rhozero > 0.9$, can be determined to lie within the regions
$\alpha > 135^{\circ}$ and $\alpha < 30^{\circ}$.  From
Fig.~\ref{fig:hM1range}, we see that $M_1/M_2 \agt 0.55$ in the gaugino
region, and therefore, even including experimental uncertainties in the
mass determinations, it is possible from measurements of $\mchargino$
and $\mLSP$ to determine that {\it either} the LSP is not a good dark
matter candidate {\it or} the grand unification prediction of $M_1/M_2
= 0.5$ is not satisfied.

The $\mslep$ bounds from $\stot$ are shown in
Fig.~\ref{fig:hmsnurange}.  The bounds in the gaugino region are again
strong, as $\stot$ is sensitive to $\mslep$ in that region.  We see
that if the underlying parameters lie in the gaugino region, the bound
$\mslep \alt 250 \gev$ applies, a promising result for scalar particle
searches.  The limits near $\alpha=90^\circ$ are not as strong, which
is hardly surprising, since in the Higgsino region $\stot$ is highly
insensitive to $\mslep$.

The bounds on $\msq$ from $B_l$ are presented in
Fig.~\ref{fig:hmsqrange}, where we have magnified two regions of
parameter space that may be excluded based on the $B_l$ measurement. As
we saw in Sec.~\ref{subsec:BrLepton}, there is generically a pocket of
small $B_l$ in the mixed $\mu <0$ region for moderate $\tanb$. In the
cross-hatched, crescent-shaped excluded region in
Fig.~\ref{fig:hmsqrange}a, $B_l$ would be much smaller than the
observed measurement of $0.34$ even for the largest values of $\msq$.
In Fig.~\ref{fig:hmsqrange}b, we see that the hatched far gaugino
region is excluded because $B_l$ would be too high, even for the lowest
allowed value of $\msq$.  For $\tanb = 1$, this excluded region is for
$\alpha \agt 177^{\circ}$ ($\mu \alt -1.5 \tev$), a region that is
already disfavored by fine-tuning considerations.  However, for larger
$\tanb$, as discussed in Sec.~\ref{subsec:BrLepton}, $B_l$ grows more
quickly as one approaches the far gaugino limit. The excluded region is
therefore larger for higher $\tanb$, and, for $\tanb = 10$, points with
$\alpha \agt 170^{\circ}$ ($\mu \alt -450 \gev$) are excluded.

Compiling these results, along with those for $M_1 < 0$, we find that
the allowed regions are as given in Fig.~\ref{fig:hallowed}.  Although
we have seen that a number of interesting correlations hold, it is
clear that the global bounds will not be as impressive as in the
gaugino case. Nevertheless, we present them here for completeness:

\be\label{hrangesall}
\begin{array}{rcccl}
0.00    &<& \rhoplus        &<& 1.00 \\
0.01    &<& \rhozero        &<& 1.00 \\
100\gev &<& \mslep          &<& 1 \tev \\
-1.00    <  \frac{M_1}{M_2}  <  -0.03
&\quad& {\rm or} &\quad&
0.10     <  \frac{M_1}{M_2}  <  0.99 \\
-870\gev <  \mu              <  -65 \gev
&\quad& {\rm or} &\quad&
79 \gev <  \mu               < 1 \tev \\
62\gev  &<& M_2             &<& 1 \tev \\
1       &<& \tanb           &<& 50, \\
150\gev &<& \msq            &<& 1\tev \ .
\end{array}
\ee

In Fig.~\ref{fig:hafb} we plot $\afbjj$ vs. $\afbchi$ for a number of
points in the allowed region.  The solid line is the measured value of
$\afbjj$, and the dashed lines are the 1$\sigma$ bounds.  Although the
points only sparsely sample the allowed region, it is evident that the
relation between the production quantity $\afbchi$ and the observed
$\afbjj$ is heavily dependent on the decay process, and in fact, for
the measured $\afbjj$, the full range of $\afbchi$ values is possible.
Without densely sampling the allowed region, it is difficult to reach
any clear conclusions about the specific shape of the regions excluded
by $\afbjj$, but it is likely that properties of the decay process will
be useful in further reducing the allowed parameter space.

\subsection{Mixed Region}

Finally, we turn to an example in the mixed region with parameters

\be\label{mparam}
(\mu, M_2, \tanb, M_1/M_2, \mslep, \msq) = (-90,115,4,0.5,200,300) \ ,
\ee
for which $\alpha = 128^\circ$ and

\be\label{mvalues}
\begin{array}{rcl}
\mchargino &=& 80.3 \gev \\
\mLSP      &=& 52.8 \gev \\
\stot      &=& 0.75 {\rm R} = 2070 \fb \\
\rhoplus   &=& 0.64 \\
\rhozero   &=& 0.60 \\
B_l        &=& 0.32 \\
m_{jj}     &<& 27.6 \gev \\
12.6 \gev  &<& E_{jj} < 41.4 \gev \ .
\end{array}
\ee
This point in parameter space has been chosen to give the same
$\mchargino$ as in the previous cases, and a value of $\rho$ that is
between those of the earlier examples. For this point, the mass
spectrum of the charginos and neutralinos has two features not present
in the previous two cases. The second neutralino has mass $m_{\chn_2} =
76.6 \gev$, and is therefore lighter than the lighter chargino. This
means that cascade decays of the chargino are kinematically possible.
However, as $\mchargino - m_{\chn_2} \gg \mchargino - \mLSP$, and
direct decays are not suppressed by any small couplings in the mixed
region, cascade decays are highly suppressed relative to direct decays
to the LSP, and we do not expect them to alter our analysis.  The
second new feature is that, since neither $M_2$ nor $\mu$ is large,
even $\chn_3$ is light with mass 119 GeV. Thus, in this case not only
are $\chn_1 \chn_2$ and $\chn_2 \chn_2$ production possible, but even
$\chn_1 \chn_3$ production is possible. The simultaneous production of
all these signals may make chargino production difficult to resolve.
However, we may remove part of this background by reducing the beam
energy below the $\chn_1 \chn_3$ production threshold. In this study we
will assume that $\chn_3$ production can be separated from chargino
production through this procedure, and we will ignore the effects of
$\chn_3$ production as a background to chargino events. Despite
possible difficulties from an entanglement of many supersymmetric
signals, it should be kept in mind that every signal brings a wealth of
new information, and generically the mixed region is the most, not the
least, optimistic scenario. Though we will consider only the
constraints that may be extracted from the chargino signal, the
neutralino signals will lead to additional restrictions that should be
imposed on the parameter space, and the full set of constraints from
LEP II will most likely be stronger than our results would suggest.

In arriving at bounds for the more interesting quantities, we will skip
many details as the method is identical to that employed in the
previous cases. Given a sample of $1 \ifb$, there are 2072 events of
which 907 are mixed mode events.  Of these, 741 are $Y$ mode events,
and 444 (60\%) of these survive the cuts of Sec.~\ref{sec:simulation}.

{}From plots of the distributions of $m_{jj}$ and $E_{jj}$, we estimate
the endpoint uncertainties to be 3 GeV for $\mjjmax$ and 2 GeV for
$\Ejjmax$.  Again assuming that tracking chamber measurements reduce
these uncertainties by 25\%, we find that $\Delta\mjjmax = 1.5 \gev$
and $\Delta\Ejjmax = 2.3 \gev$. The resulting uncertainties for
$\mchargino$ and $\mLSP$ are

\be
\Delta\mchargino = 3.3 \gev \quad {\rm and} \quad
\Delta\mLSP  = 2.7 \gev \ .
\ee

To determine the uncertainties in the determinations of $\stot$ and
$B_l$, we must first determine $\Delta\sigy / \sigy$ from
Eq.~(\ref{sigmai}). With 444 $Y$ mode events, the statistical
uncertainty is $\Delta N_Y/ N_Y = 4.7\%$.  The efficiencies of the cuts
for a wide range of representative points in SUSY parameter space,
subject only to the restriction $\mchargino \approx 80\gev$ and $\mLSP
\approx 53\gev$, range from 54--62\%, and we therefore take the
efficiency to be $\eta_Y = 58\%$ and its variation to be $\Delta\eta_Y
= 4\%$.  Combining these uncertainties, we find that $\Delta\sigy
/\sigy = 8.4\%$. There are 975 Monte Carlo hadronic mode events before
cuts, and again taking $\eta$ and $\Delta\eta$ to be approximately
equal for the $Y$ and hadronic modes, we find $\Delta\sigh /\sigh =
8.1\%$.  Using Eq.~(\ref{stotblerr}) we determine that

\be
\Delta\stot /\stot = 6.1\% \quad {\rm and} \quad \Delta B_l /B_l =
7.9\% \ .
\ee

Given these ranges for the four observables, we may now bound the
parameter space. The allowed range of $M_1/M_2$ is shown in
Fig.~\ref{fig:mM1range}. In the gaugino region, the minimum value of
$M_1/M_2$ is roughly 0.5; though values of $M_1/M_2>0.5$ are favored,
the prediction of grand unified theories cannot be excluded. As in the
previous two examples, a small $\mu<0$, $\tanb\approx 1$ region is
ruled out.

The $\mslep$ bounds from $\stot$ are shown in
Fig.~\ref{fig:mmsnurange}.  For this example $\stot$ lies {\it below}
the value of $\stot$ approached in the pure Higgsino limit, as may be
seen from Eq.~(\ref{stotH}) or Fig.~\ref{fig:sigmasnu}.  Thus, not only
is some of the Higgsino region excluded, but also we obtain, from the
$\stot$ measurement alone, an upper bound on $\mslep$. For $\tanb>4$ we
see that the low value of $\stot$ gives the bound $\mslep<250 \gev$.

Because $B_l$ is approximately $\frac{1}{3}$ as it was in the Higgsino
example, the bounds on $\msq$ are fairly similar to those obtained in
Fig.~\ref{fig:hmsqrange}, so we will not present the $(\msq)_{\rm min}$
and $(\msq)_{\rm max}$ contours for this case study. It is again
possible to rule out a crescent-shaped region in which $B_l$ is too
small, and the far gaugino region in which $B_l$ is too large.  The
allowed region for the mixed region case study, considering both
negative and positive $M_1$, is given in Fig.~\ref{fig:mallowed}. By
randomly sampling the allowed region, we find the following global
bounds:

\be\label{mrangesall}
\begin{array}{rcccl}
0.05    &<& \rhoplus        &<& 1.00 \\
0.01    &<& \rhozero        &<& 1.00 \\
100\gev &<& \mslep          &<& 257 \gev \\
-1.00    <  \frac{M_1}{M_2}  <  -0.12
&\quad& {\rm or} &\quad&
0.16     <  \frac{M_1}{M_2}  <  1.00 \\
-339\gev <  \mu              <  -52 \gev
&\quad& {\rm or} &\quad&
85 \gev <  \mu               < 355 \gev \\
51\gev  &<& M_2             &<& 500 \gev \\
1     &<& \tanb           &<& 50, \\
150\gev &<& \msq            &<& 1 \tev \ .
\end{array}
\ee
Again, the correlations among the various parameters are not
represented in these limits.  As already noted from
Fig.~\ref{fig:mmsnurange}, we see that the sneutrino mass bound is very
stringent, with a maximum value of 257 GeV, as in the gaugino case. The
other global bounds are weak.

Finally, we plot $\afbjj$ vs. $\afbchi$ for a few points in the allowed
region in Fig.~\ref{fig:mafb}.  As in the previous figures, the solid
line is the measured value of $\afbjj$, and the dashed lines are the
1$\sigma$ bounds.  Although definite conclusions would require a more
thorough sampling of the allowed region, Fig.~\ref{fig:mafb} suggests
that the point we have picked has an extreme value of $\afbjj$, and
could therefore be distinguished from most other points in the allowed
region by decay process considerations.

\section{Final Comments and Summary}
\label{sec:conc}

We have explored the potential for precise determinations of
fundamental SUSY parameters from chargino production at LEP II. We have
found that chargino events can be well-described by six underlying SUSY
parameters: $\mu, M_2, \tanb, M_1, \mslep$, and $\msq$. A number of
observables were investigated, and four --- the chargino mass, the LSP
mass, the total cross section, and the leptonic branching fraction ---
were found to be particularly useful in most areas of parameter space.
These four observables, with their accompanying uncertainties, were
used to restrict the allowed SUSY parameter space for representative
points in the gaugino, Higgsino, and mixed regions, and a simple method
for representing the results graphically was used.

In the gaugino region, we found stringent global bounds on SUSY
parameters.  In particular, $\rhozero$ was restricted to ranges in
which the LSP is a good dark matter candidate, the gaugino mass
unification condition could be verified or disproved at the level of
15\%, and an upper limit for the sneutrino mass could be obtained for
an underlying value of $\msnu = 200 \gev$. We also found that the
results for $\rhozero$ and $\msnu$ were not altered substantially when
significantly worse experimental conditions were assumed. In the
Higgsino case study, stringent {\it global} bounds were not found for
any of the combinations of SUSY parameters.  However, a number of
interesting correlations were found, making it possible, for example,
to exclude the grand unification condition $M_1/M_2 = 0.5$ in the
gaugino region.  In the mixed region example, results similar to those
for the Higgsino region were achieved, with the exception that it was
once again possible to set a stringent global upper bound on $\msnu$.

In this study, we have only crudely simulated chargino events and
detector effects.  Although we have shown that, at least in some cases,
our results are not very sensitive to the exact experimental
assumptions, detailed event simulations and detector modeling would
sharpen our results.  Other work that may improve the results obtained
here includes a study of chargino production at threshold
\cite{decaystudy}, where chargino decays are more easily analyzed, and
investigations of other SUSY processes, notably neutralino production,
which may provide useful constraints in some regions of SUSY parameter
space.

In summary, our results imply that if charginos are discovered at LEP
II, they will bring not only the first experimental evidence for SUSY,
but will also significantly restrict SUSY parameter space and may
provide bounds on SUSY parameters of relevance to cosmology, grand
unified theories, and future sparticle searches.

\acknowledgements

It is a pleasure to thank M.~Peskin for valuable comments and
suggestions throughout the course of this work.  In addition, we are
grateful to T.~Barklow for many discussions on experimental issues, and
to Y.~Kizukuri and N.~Oshimo for their generosity in helping us correct
an error in our calculation of chargino branching fractions.  We also
thank L.~Dixon, G.~Farrar, M.~Felcini, H.~E.~Haber, and H.~Murayama for
helpful conversations.

\appendix*{Uncertainty Analysis}

In this appendix we collect the various formulae for calculating
uncertainties needed in the preceding sections.

The chargino and LSP masses determine the maximum and minimum dijet
energies, $\Ejjmax$ and $\Ejjmin$, and the maximum dijet mass,
$\mjjmax$.  Thus, if two of these three endpoints are measured, one can
determine $\mchargino$ and $\mLSP$.

In terms of $\mjjmax$ and either one of the energy endpoints $E_{jj}^0
\equiv \Ejjmax$ or $\Ejjmin$, the masses $\mchargino$ and $\mLSP$ are
given by

\be\label{massesfrommE}
\begin{array}{rcl}
\mchargino (\mjjmax, \Ejjm ) &=& \frac{\ds\mjjmax}{\ds 2}\frac{\ds 2
\Ejjm E_b
+ (\mjjmax) ^2 + \Ejjm\sqrt{4 E_b^2-4\Ejjm E_b-m^2}}{\ds
(\Ejjm)^2 + (\mjjmax)^2}\ , \\
\\
\mLSP (\mjjmax, \Ejjm ) &=& \mchargino (\mjjmax, \Ejjm ) - \mjjmax
\ ,
\end{array}
\ee
where $E_b$ is the beam energy.  The uncertainties in the mass
determinations are given simply by adding the endpoint uncertainties in
quadrature:

\be\label{masserror1}
\Delta\mchargino=
\left[ \left( \frac{\ds\partial\mchargino}{\ds\partial \mjjmax}
\Delta\mjjmax \right)^2 +
\left( \frac{\ds\partial\mchargino}{\ds\partial \Ejjm}
\Delta\Ejjm \right)^2 \right] ^{\half} \ ,
\ee
and similarly for $\Delta\mLSP$.

To calculate $\stot$ and $B_l$, one must measure at least two partial
cross sections.  If one has measured the mixed and hadronic partial
cross sections $\sigm$ and $\sigh$, the total cross section and
leptonic branching fraction are

\be
\begin{array}{rcl}
\stot &=& \frac{\ds (\sigm+2\sigh)^2}{\ds 4\sigh} \ , \\
\\
B_l   &=& \frac{\ds \sigm}{\ds \sigm+2\sigh} \ ,
\end{array}
\ee
and their fractional uncertainties are given by

\be\label{stotblerr}
\begin{array}{rcl}
\left(\frac{\ds \Delta\stot}{\ds \stot}\right)^2
&=& 4 B_l^2 \left(\frac{\ds \Delta\sigm}{\ds \sigm}\right)^2
  + (1-2 B_l)^2
  \left(\frac{\ds \Delta\sigh}{\ds \sigh}\right)^2 \ , \\
\\
\left(\frac{\ds \Delta B_l}{\ds B_l}\right)^2
&=& (1-B_l)^2 \left[\left(\frac{\ds \Delta\sigm}{\ds \sigm}\right)^2
  + \left(\frac{\ds \Delta\sigh}{\ds \sigh}\right)^2 \right] \ .
\end{array}
\ee
If $\sigm$ and $\sigl$ are measured instead,

\be\label{stotblerr2}
\begin{array}{rcl}
\stot &=& \frac{\ds (\sigm+2\sigl)^2}{\ds 4\sigl} \ , \\
\\
B_l   &=& \frac{\ds 2\sigl}{\ds 2\sigl+\sigm} \ ,
\end{array}
\ee
and

\be
\begin{array}{rcl}
\left(\frac{\ds \Delta\stot}{\ds \stot}\right)^2
&=& 4 (1-B_l)^2 \left(\frac{\ds \Delta\sigm}{\ds \sigm}\right)^2
  + (1-2 B_l)^2
  \left(\frac{\ds \Delta\sigl}{\ds \sigl}\right)^2 \ , \\
\\
\left(\frac{\ds \Delta B_l}{\ds B_l}\right)^2
&=& (1-B_l)^2 \left[\left(\frac{\ds \Delta\sigm}{\ds \sigm}\right)^2
  + \left(\frac{\ds \Delta\sigl}{\ds \sigl}\right)^2 \right]  \ .
\end{array}
\ee
If all three partial cross sections are measured, they may all be used
to improve the determinations of $\stot$ and $B_l$. In this study, the
$Y$ mode partial cross section is measured instead of the mixed mode.
After lepton universality is verified, $\sigm = \frac{5}{4} \sigy$, and
one can simply replace $\sigm$ by $\sigy$ in the above formulae.

Finally, we must calculate the uncertainty of $\afbjj$. The
forward-backward asymmetry is $\afbjj = 2p-1$, where $p$ is the
fraction of events in the forward hemisphere, and the standard error of
an estimate of a population proportion is

\be
\sigma (p) = \sqrt{\frac{p(1-p)}{N}} \ ,
\ee
where $N$ is the number of samples.  The 1$\sigma$ uncertainty in the
measurement of $\afbjj$ is therefore given by

\be\label{afberror}
\sigma (\afbjj ) = 2 \sigma (p) = \sqrt{\frac{1-(\afbjj) ^2}{N}} \ .
\ee

\figure{\label{fig:feynprod}
The three chargino production diagrams for \epem\ collisions.}

\figure{\label{fig:feyndecay}
The three-body chargino decay diagrams.  There exist separate scalar
partners for each chirality of fermion, and so there are a total of six
hadronic and five leptonic decay diagrams.}

\figure{\label{fig:chargcont}
Contours of constant $\mchargino$ (in GeV) for $\tanb=4$.  The
cross-hatched region is excluded by the experimental bounds $\mchargino
> 45 \gev$ and $\mLSP > 20 \gev$. In the hatched regions, $\mchargino >
95 \gev$, so charginos are kinematically inaccessible.}

\figure{\label{fig:neutcont}
Contours of constant $\mLSP$ (in GeV) for fixed $\tanb=4$ and three
values of the gaugino mass ratio $M_1/M_2$: (a) 0.5, (b) 0.75, and (c)
$-0.5$. The hatched and cross-hatched regions are as in
Fig.~\ref{fig:chargcont}.}

\figure{\label{fig:rhocont}
Contours of constant $\rhoplus$ for $\tanb = 4$, with hatched and
cross-hatched regions as in Fig.~\ref{fig:chargcont}. As defined in the
text, $0 \le \rhoplus \le 0.2$ in the Higgsino region, $0.2< \rhoplus
<0.9$ in the mixed region, and $0.9\le \rhoplus \le 1.0$ in the gaugino
region. The gaugino content approaches one asymptotically as $|\mu|
\rightarrow \infty$ in the gaugino region and zero asymptotically as
$M_2 \rightarrow \infty$ in the Higgsino region.}

\figure{\label{fig:emax}
Contours of constant $E^{\rm max}$ (in GeV), the maximum jet or lepton
energy for decay products from chargino events.  The plot is for
$M_1/M_2=0.5$ and $\tanb=4$. The hatched and cross-hatched regions are
as in Fig.~\ref{fig:chargcont}, and the vertical and horizontal scales
are chosen to emphasize the Higgsino region.  We see that in the
Higgsino region, for points in parameter space with large $M_2$ or near
threshold, the decay products may be too soft to be useful for
precision measurements. }

\figure{\label{fig:threshold}
The total cross section $\stot$ before cuts (solid curve) as a function
of $\sqrt{s}$ for parameters  $(\mu, M_2, \tanb, \msnu) = (-400, 75, 4,
200)$ in the gaugino region. For comparison, a unit of R is also
plotted (dashed curve). We note the sudden rise in cross section,
characteristic of fermion production.}

\figure{\label{fig:Ac}
Contours of constant value of the ratio $A_{{\rm central}}^{\chc}/
A_{{\rm central}}^{(0)\chc}$, defined in the text. The hatched and
cross-hatched regions are as in Fig.~\ref{fig:chargcont}. The contours
are plotted in the $\mum$ plane for fixed $\tanb=4$ and (a) $\msnu=200
\gev$ and (b) $\msnu=100 \gev$ and are chosen to emphasize the range of
the ratio. In the case $\msnu = 200 \gev$, where the perturbative
analysis is expected to hold, we see that the ratio is approximately
one, as expected. For $\msnu=100\gev$, beyond the range in which the
perturbative analysis can be expected to be valid, we see that this
behavior nevertheless persists, and deviations are small.}

\figure{\label{fig:sigma}
Contours of constant $\stot$ in picobarns in the $\mum$ plane for fixed
$\tanb=4$ and (a) $\msnu = 1 \tev$ and (b) $\msnu = 150\gev$.  The
hatched and cross-hatched regions are as in Fig.~\ref{fig:chargcont}.
The cross section rises quickly near threshold. The lower cross
sections in (b) are a result of the large destructive interference of
the sneutrino diagram.}

\figure{\label{fig:sigmasnu}
Cross sections as functions of $\msnu$ are plotted for $\tanb=4$ and
four representative $\mum$ points: $(-400, 75)$ in the gaugino region
(solid), $(-90, 115)$ in the mixed region (dashed), $(-75, 250)$ in the
Higgsino region (dot-dashed), and $(-78, 1000)$ in the far Higgsino
region (dotted). The $\snu$ diagram gives a large and destructive
contribution for the gaugino and mixed points. As the Higgsino
component increases, the dependence of the cross section on $\msnu$
decreases.}

\figure{\label{fig:afbfig}
Contours of constant $\afbchi$ in percent for $\tanb=4$ and (a) $\msnu
= 1 \tev$ and (b) $\msnu = 150\gev$. The hatched and cross-hatched
regions are as in Fig.~\ref{fig:chargcont}. In the large $\msnu$ limit,
$\afbchi \rightarrow 0$ in both the gaugino and Higgsino limits, with
only the mixed region giving a significant negative $\afbchi$.  For low
$\msnu$, the $t$-channel sneutrino exchange diagram can lead to large
and positive $\afbchi$.}

\figure{\label{fig:sigmalr}
Contours of constant $\sigma_R/\sigma_L$ in percent in the $\mum$ plane
for $\tanb=4$ and $\msnu=150 \gev$.  The ratio never rises above 15\%
in the allowed bands, and is approximately 2\% in much of the gaugino
region. The hatched and cross-hatched regions are as in
Fig.~\ref{fig:chargcont}.}

\figure{\label{fig:Bl}
Contours of constant value of the leptonic branching fraction $B_l$ in
the $\mum$ plane for $M_1/M_2=0.5$ and three sets of parameters
$(\tanb, \mslep, \msq)$: (a) (2, 200, 200), (b) (2, 200, 800), and (c)
(10, 200, 200). The gaugino region has been magnified, and the hatched
and cross-hatched regions are as in Fig.~\ref{fig:chargcont}. For all
figures, the value of $B_l$ is $\frac{1}{3}$ in the Higgsino region and
grows as one approaches the gaugino region.  The growth is faster for
large $\tanb$ (c) than for low $\tanb$ (a). In (a) and (b) the $B_l$
contours differ by approximately 5\% in the far gaugino region,
consistent with Eq.~(\ref{lhratioisospin}). Note also the ``pocket'' in
the $\mu<0$ near gaugino region, where $B_l < \frac{1}{3}$. }

\figure{\label{fig:sheets}
The $\mchargino$ measurement restricts the $(\mu, M_2, \tanb)$ space to
two thin sheets $\cal S$, which are then flattened into the plane $\cal
T$ with the transformation $(\mu, M_2, \tanb) \rightarrow (\alpha,
\tanb)$, where $\alpha = \arctan (M_2/\mu)$. This transformation is
illustrated schematically here.  For large $\tanb$, observables are
symmetric under $\mu\leftrightarrow -\mu$, that is, under
$\alpha\leftrightarrow 180^\circ-\alpha$.}

\figure{\label{fig:gmjjEjj}
The dijet (a) mass spectrum and (b) energy spectrum, after cuts, for
the gaugino case $(\mu, M_2, \tanb, M_1/M_2, \mslep, \msq)$ =
$(-400,75,4,0.5,200,300)$ with integrated luminosity $1 \ifb$. In these
distributions, hadrons from $\tau$ lepton decays have not been
included. We see that finite detector resolution effects cause the
spectra to have tails that exceed the theoretical limits, but despite
this, the endpoints are fairly sharp. We estimate that the 1$\sigma$
uncertainty of $\mjjmax$ is 2 GeV, and that for $\Ejjmax$ is 3 GeV.
Note that very few events have dijets with low invariant mass, and it
is therefore possible to distinguish hadrons that result from $\tau$
decays and those that result from hadronic chargino decays.}

\figure{\label{fig:gM1range}
Gaugino example contours for the (a) minimum and (b) maximum values of
$M_1/M_2$ in the $(\alpha, \tanb )$ plane, as defined in
Sec.~\ref{sec:strategy}. The cross-hatched area in the $\alpha >
90^{\circ}$ mixed region with $\tanb\approx 1$ is excluded by the
$\mLSP$ measurement. The approximate symmetry $\alpha \leftrightarrow
180^{\circ}-\alpha$ ($\mu \leftrightarrow -\mu$) for large $\tanb$ is
already in evidence at $\tanb = 10$. }

\figure{\label{fig:gmsnurange}
Gaugino example contours for the (a) minimum and (b) maximum values of
$\mslep$ in the $(\alpha, \tanb)$ plane, as defined in
Sec.~\ref{sec:strategy}.  The cross-hatched area in the Higgsino region
is excluded because the measured $\stot$ is too large to be compatible
with any $\mslep$. }

\figure{\label{fig:gmsqrange}
Gaugino example contours for the (a) minimum and (b) maximum values of
$\msq$ in the $(\alpha, \tanb )$ plane, as defined in
Sec.~\ref{sec:strategy}.  The $\alpha> 90^{\circ}$ ($\mu<0$) gaugino
region has been magnified. (Similar features are seen in the $\alpha <
90^{\circ}$ ($\mu > 0$) gaugino region.) The hatched and cross-hatched
regions are excluded by the $B_l$ measurement: in the hatched region
$B_l$ is too large, and in the cross-hatched region $B_l$ is too small.}

\figure{\label{fig:gallowed}
The allowed region in the $(\alpha, \tanb )$ plane for the gaugino case
study with both signs of $M_1$ allowed. The hatched regions are
excluded by the measurements of $\mchargino$, $\mLSP$, $\stot$, and
$B_l$, which confine the allowed region to narrow strips in the gaugino
region. The dot indicates the underlying value of $(\alpha, \tanb )$
for the gaugino case study. }

\figure{\label{fig:gafb}
Plot of $\afbjj$ vs. $\afbchi$ for several points in the allowed region
of the gaugino case study.  The solid line is the measured $\afbjj$
from the Monte Carlo simulation, and the dashed lines give the
1$\sigma$ uncertainties in this measurement.  Points with $\mu <0$ are
given by filled circles, and all points with $\mu>0$ (open circles) are
seen to be excluded by the $\afbjj$ measurement.  The case study is
indicated by a star.}

\figure{\label{fig:hmjjEjj}
The dijet (a) mass spectrum and (b) energy spectrum, after cuts, for
the Higgsino case $(\mu, M_2, \tanb, M_1/M_2, \mslep, \msq)$ =
$(-75,250,4,0.5,200,300)$ with integrated luminosity $1 \ifb$. We
estimate that the 1$\sigma$ uncertainty of $\mjjmax$ is 1 GeV, and that
for $\Ejjmax$ is 2 GeV. }

\figure{\label{fig:hM1range}
Higgsino example contours for the (a) minimum and (b) maximum values of
$M_1/M_2$. In (b), two $M_1/M_2=0.8$ contours have been removed from
the Higgsino region for clarity. The cross-hatched area in the $\alpha>
90^{\circ}$ mixed region with $\tanb\approx 1$ is excluded by the
$\mLSP$ measurement alone. }

\figure{\label{fig:hmsnurange}
Higgsino example contours for the (a) minimum and (b) maximum values of
$\mslep$.  Stringent bounds are found in the gaugino region, but no
limits are obtained in the Higgsino region. }

\figure{\label{fig:hmsqrange}
Higgsino example contours for the (a) minimum and (b) maximum values of
$\msq$. Two areas in the $\alpha> 90^{\circ}$ ($\mu<0$) gaugino region
has been magnified. The cross-hatched region of (a) and the hatched
region of (b) have $B_l$ values that are too low and too high,
respectively, and are excluded.}

\figure{\label{fig:hallowed}
The allowed region in the $(\alpha, \tanb )$ plane for the Higgsino
case study, with both signs of $M_1$ allowed. The cross-hatched regions
are excluded by the measurements of $\mchargino$, $\mLSP$, $\stot$, and
$B_l$.  The dot indicates the value of $(\alpha, \tanb )$ for the case
study.}

\figure{\label{fig:hafb}
Plot of $\afbjj$ vs. $\afbchi$ for several points in the allowed region
of the Higgsino case study.  The solid line is the measured $\afbjj$
from the Monte Carlo simulation, and the dashed lines give the
1$\sigma$ uncertainties in this measurement. The case study is
indicated by a star. }

\figure{\label{fig:mM1range}
Mixed example contours for the (a) minimum and (b) maximum values of
$M_1/M_2$. The cross-hatched area in the mixed region with
$\tanb\approx 1$ is excluded by the $\mLSP$ measurement alone. }

\figure{\label{fig:mmsnurange}
Mixed example contours for the (a) minimum and (b) maximum values of
$\mslep$.  The cross-hatched area in the Higgsino region has $\stot$
values that are higher than that measured and is excluded. In this
case, an upper bound on $\mslep$ may be obtained from the $\stot$
measurement alone. }

\figure{\label{fig:mallowed}
The allowed region in the $(\alpha, \tanb )$ plane for the mixed region
case study with both signs of $M_1$ allowed. The cross-hatched regions
are excluded by the measurements of $\mchargino$, $\mLSP$, $\stot$, and
$B_l$, and the dot indicates the value of $(\alpha, \tanb )$ for the
case study.}

\figure{\label{fig:mafb}
Plot of $\afbjj$ vs. $\afbchi$ for several points in the allowed region
of the mixed region case study.  The solid line is the measured
$\afbjj$ from the Monte Carlo simulation, and the dashed lines give the
1$\sigma$ uncertainties in this measurement.  The specific parameters
chosen for the case study lie at an extreme value of $\afbjj$ and
indicate that this measurement may be able to reduce the allowed region
substantially. The case study is indicated by a star. }

\end{document}